\documentclass{article}
\usepackage{graphicx}
\usepackage[T1]{fontenc}
\usepackage{textcomp}
\usepackage{natbib}
\usepackage{epsfig}
\usepackage{color}
\usepackage[english]{babel}
\usepackage{enumerate}
\usepackage{amsmath}
\usepackage{subfigure}
\usepackage{multirow}
\usepackage{appendix}
\usepackage[margin=0.8in]{geometry}

\newcommand{\btau}{\mbox{\boldmath{$\tau$}}}

\newcommand{\bmu}{\mbox{\boldmath{$\mu$}}}
\newcommand{\btheta}{\mbox{\boldmath{$\theta$}}}

\newcommand{\bSigma}{\mbox{\boldmath{$\Sigma$}}}

\begin{document}

\begin{center}

\huge{Investigation of Parameter Uncertainty in Clustering Using a Gaussian Mixture Model Via Jackknife, Bootstrap and Weighted Likelihood Bootstrap.}

\vspace{10mm}

\large{This work is supported by the Insight Research Centre (SFI/12/RC/2289) and Science Foundation Ireland under the Research Frontiers Programme (2007/RFP/MATH281)}


\vspace{10mm}
Adrian O'Hagan$^1$,\,\,
\vspace{2mm}
Thomas Brendan Murphy$^2$,\,\,
\vspace{2mm}
Luca Scrucca$^3$\,\,
\vspace{2mm}
\&
Isobel Claire Gormley$^4$


\vspace{10mm}
$^1$ School of Mathematics and Statistics \& Insight: Centre for Data Analytics, University College Dublin, Ireland.

adrian.ohagan@ucd.ie
              
\vspace{3mm}
$^2$ School of Mathematics and Statistics \& Insight: Centre for Data Analytics, University College Dublin, Ireland
                 
\vspace{3mm}     
$^3$ Department of Economics, Universit\`a degli Studi di Perugia, Italy

\vspace{3mm}
$^4$ School of Mathematics and Statistics \& Insight: Centre for Data Analytics, University College Dublin, Ireland.

\end{center}

\newpage
\section{Abstract}
Mixture models with (multivariate) Gaussian components are a popular tool in model-based clustering. Such
models are often fitted by a procedure that maximizes the likelihood, such as the EM algorithm. At convergence, the
maximum likelihood parameter estimates are typically reported, but in most cases little emphasis is placed on the variability
associated with these estimates. In part this may be due to the fact that standard errors are not directly calculated in the model-fitting
algorithm, either because they are not required to fit the model, or because they are difficult to compute. The examination of standard errors in model-based clustering is therefore typically neglected.

Sampling based methods, such as the jackknife (JK), bootstrap (BS) and parametric bootstrap (PB), are intuitive, generalizable approaches to assessing parameter uncertainty in model-based clustering using a Gaussian mixture model. This paper provides a review and empirical comparison of the jackknife, bootstrap and parametric bootstrap methods for producing standard errors and confidence intervals for mixture parameters. The performance of such sampling methods in the presence of small and/or overlapping clusters requires consideration however; here the weighted likelihood bootstrap (WLBS) approach is demonstrated to be effective in addressing this concern in a model-based clustering framework. The JK, BS, PB and WLBS methods are illustrated and contrasted through simulation studies and through the traditional \emph{Old Faithful} data set and also the \emph{Thyroid} data set. The \texttt{MclustBootstrap} function, available in the most recent release of the popular \texttt{R} package \textbf{mclust}, facilitates the implementation of the JK, BS, PB and WLBS approaches to estimating parameter uncertainty in the context of model-based clustering.

The JK, WLBS and PB approaches to variance estimation are shown to be robust and provide good coverage across a range of real and simulated data sets when performing model-based clustering; but care is advised when using the BS in such settings. In the case of poor model fit (for example for data with small and/or overlapping clusters), JK and BS are found to suffer from not being able to fit the specified model in many of the sub-samples formed. The PB also suffers when model fit is poor since it is reliant on data sets simulated from the model upon which to base the variance estimation calculations. However the WLBS will generally provide a robust solution, driven by the fact that all observations are represented with some weight in each of the sub-samples formed under this approach.

\vspace{3mm}
keywords: \texttt{mclust}, \texttt{MclustBootstrap}, Precision, Standard Errors, Variance Estimation.

\newpage

\section{Introduction}
\label{sec:intro}

The bootstrap (BS) and jackknife (JK) methods of re-sampling originated as a
non-parametric means of estimating the variability of parameter
estimates, or of estimating the parameters themselves
\citep{quenouille56, tukey58}. These methods have since been
documented and studied in great detail, in a wide range of modeling
scenarios, such as regression \citep{wu1986jackknife},
generalized linear models \citep{moulton1991bootstrapping} and time series analysis \citep{buhlmann1997sieve}.
The literature includes applications of the jackknife or
bootstrap in isolation (see for example \cite{efron81} and
\cite{Efron:Tibshirani:1993} respectively) as well as applications of both
methods together, exploring the synergy between them
\citep{efron81_2}. This synergy encompasses the fact that
the methods are asymptotically equivalent, as well as the fact that
both methods derive from the same basic idea.
Therefore, while the algorithms for the two methods have
traditionally been presented separately, it is important to note their
shared objective: constructing new samples from the original data that allow us to
gauge the variability of parameter estimates for the estimated model.

The methods can be helpful either in cases where the robustness of
the parametric assumptions is in question, or when the second
moments of the sampling distribution of the parameter estimates are
difficult to compute. The former problem often arises when analyzing
``small'' data sets. The latter issue is often present in more difficult model settings,
such as model-based clustering, where the log-likelihood can become
intractable for complex distributions. Indeed this may help to
explain the lack of attention paid to the variability of point
estimates under this approach to statistical modeling, compared to
other standard methods such as regression or ANOVA.

The JK, BS and PB methods provide fast and accurate ways to
circumvent such problems and generate variance estimates for maximum likelihood parameter estimates. This paper provides a review and empirical comparison of the JK, BS and PB methods for producing standard errors and confidence intervals for mixture parameters in the context of model-based clustering with multivariate Gaussian components. The performance of such sampling methods in the presence of small and/or overlapping clusters requires consideration however; here, the weighted likelihood bootstrap (WLBS) approach is demonstrated to be effective in addressing this concern.
The procedures are illustrated when clustering using a mixture of Gaussian
distributions in simulation studies and in the case of the \emph{Old Faithful} and \emph{Thyroid} data, well-documented multivariate clustering test cases. In particular, the methods are considered within the context of the widely used \textbf{R} package \texttt{mclust} \citep{R2017, fraley02, fraley12}, which facilitates model-based clustering by considering a range of parsimonious mixtures of Gaussian distributions.
The JK, BS, PB and WLBS methods presented here are easily implemented in the most recent release of the \texttt{mclust} package, through the \texttt{MclustBootstrap} function, for which sample code is provided.

The remainder of the paper is structured as follows. Sections $\ref{sec:methods}$ provides a unified summary of the JK, BS and PB methods of variance estimation. The similarities and differences between the approaches are identified as they arise, in Sections $\ref{sec:jkbs}$ and $\ref{sec:PB}$. The motivation for and the details of the WLBS are discussed in Section $\ref{sec:WLBS}$. Section $\ref{sec:data}$ presents the illustrative data sets used -- the data sets employed in the simulation studies and the \emph{Old Faithful} and \emph{Thyroid} real data sets. In Section $\ref{sec:results}$ the results obtained for the illustrative data sets are presented and the article concludes in Section $\ref{sec:discussion}$ with a discussion of variance estimation in model-based clustering. Appendix A contains pairs plots for the variables from one of the simulated data sets tested and Appendix B contains additional parameter estimates and standard errors for the \emph{Thyroid} data set along with the \textbf{R} code used to generate them.

\vskip14pt
\section{Sampling based approaches to variance estimation in model-based clustering}
\label{sec:methods}

In a model-based clustering setting each $p$-dimensional multivariate observation
$x_i$ belongs to one of $G$ clusters. The matrix of $i = 1, \ldots, n$ observations is denoted by $\mathbf{X} = (x_1,x_2,\ldots, x_n)$. The parameter $\tau_g$ is the probability that the observation comes from cluster $g$, where $\sum_{g=1}^{G} \tau_g = 1$. The data within group $g$ are modeled by component density
$f(x_i|\btheta_g)$. For a mixture of Gaussian distributions, $\btheta_g$ comprises of the cluster means $\mu_g$ and the covariance matrices $\Sigma_g$. The observed data likelihood is the function to be maximized, however calculating maximum likelihood estimates is achieved more easily using the expected complete data likelihood. A
classification vector $z_i = (z_{i1},\ldots,z_{iG})$ is assumed to exist for each observation $i = 1,\ldots,n$ where $z_{ig} = 1$ if observation $i$ belongs to cluster $g$ and $z_{ig} = 0$ otherwise. The complete data likelihood under a finite mixture of Gaussians can be expressed as:

\begin{equation}
L_c = \displaystyle\prod_{i=1}^n\displaystyle\prod_{g=1}^G\left[\tau_g f(x_i| \mu_g, \Sigma_g)\right]^{z_{ig}}
\label{eq:complete data likelihood}
\end{equation}

The observed data likelihood is maximized via the widely used EM algorithm \citep{dempster77}, which iteratively maximises the expected complete-data log-likelihood. Extensive literature is available on fitting a Gaussian mixture model with a variety of eigendecomposed
covariance structures via the EM algorithm \citep{fraley98b, fraley02}, as is considered within the popular \textbf{R} package \texttt{mclust}. It must be noted that \texttt{mclust} provides only a local optimum of the likelihood, not a global optimum, with accompanying parameter estimates. Hence the resampling methods and parameter standard error estimation techniques detailed must be considered in this context.

While the EM algorithm can provide an efficient means of parameter estimation in the mixture modeling context, the default output of the EM algorithm does not provide estimates of the uncertainty associated with the parameter estimates. Several approaches have been considered to facilitate the provision of standard errors within the context of the EM algorithm; \cite{mclachlan97} and \cite{mclachlan00} provide thorough reviews. Most existing methods for assessing the covariance matrix of MLEs obtained via the EM algorithm are based on estimating the observed information matrix \citep{meilijson89, mclachlan97, meng89, meng91}. However, while estimating the covariance matrix of the MLEs via the information matrix is valid asymptotically \citep{boldea2009}, in the case of mixture models large sample sizes are required for the asymptotics to give a reasonable approximation. \cite{efron94} highlights that standard errors are likely to be underestimated under such approaches. Also, none of the existing information matrix based
approaches are
generalisable in that model specific alterations to the EM algorithm are required. The mixture of Gaussians approach to model-based clustering typically features non-differentiable orthogonal matrices for some covariance structures, which prohibits implementation of an information matrix-based approach. Furthermore, in certain (typically high dimensional) settings, use of the information matrix is infeasible due to singularity issues \citep{ford1980sequentially, stoica1982non, titterington1984recursive}.

Sampling based approaches promise an alternative, fast and generalisable approach to providing standard errors. Such methods are detailed in the literature: \cite{diebolt96} employ a conditional bootstrap approach to MLE covariance estimation; the EMMIX software by \cite{mclachlan99} offers parametric and nonparametric bootstrap facilities for standard error estimation; \cite{turner2000estimating} discusses non/semiparametric bootstrapping for obtaining the standard errors in a mixture of linear regressions problem as well as estimation of the observed information matrix in this setting; \cite{basford97} and \cite{peel98} compare bootstrap and information matrix approaches for Gaussian mixtures;  and \cite{nyamundanda2010} employ the jackknife for standard error estimation in the context of mixtures of constrained factor models. In a related area, \cite{mclachlan87} avails of the bootstrap to aid model selection when clustering. Here, the JK, BS and PB sampling methods, within the context of the well utilised \textbf{R} package \texttt{mclust}, are reviewed and empirically compared. Their potentially poor performance in the presence of small clusters is effectively addressed through the introduction of a weighted likelihood bootstrap (WLBS) approach.

\subsection{The jackknife and bootstrap methods}
\label{sec:jkbs}

The jackknife and bootstrap methods are well known approaches to obtaining estimates of the variance associated with parameter estimates. Both are sampling based methods and are straightforward to implement, regardless of the model under consideration. Here, they are considered within the context of model-based clustering. By default, \texttt{mclust} clusters observations by fitting a range of mixture of Gaussian models (in terms of number of mixture components and the type of covariance structure), and chooses the optimal model using the Bayesian Information Criterion \citep{schwarz1978}. However, the user may specify any covariance structure and number of groups $G$ that they wish to fit. In this setting, the algorithm for the bootstrap and jackknife variance estimation techniques proceeds as follows:

\begin{enumerate}[(i)]
\item
Identify the optimal model structure for the full data set $\mathbf{X}$, denoted by $\tilde{M}$, using \texttt{mclust}. This model provides the number of groups, $G$, and the maximum likelihood posterior group membership probability matrix $\mathbf{\hat{Z}}_{\tilde{M}}$. The value $\hat{z}_{ig}$ is the posterior probability that observation $i$ belongs to group $g$.
Note that this step has not been carried out in the subsequent simulations, rather the true model has been assumed to be known.

\item
\label{alg:sample}
Form $B$ samples comprising of observations from the original data
$\mathbf{X}$.

\begin{itemize}
 \item  Under the JK approach, each of the
$B_{JK} = n$ samples contains $(n-1)$ observations. Jackknife sample
$\mathbf{X}_j$ denotes the sample of the original observations
$\mathbf{X}$ with observation $j$ omitted, $j = 1,2,\ldots,n$.

\item Under the BS, each of the $B_{BS}$
samples contains $n$ observations, where the observations are
sampled with replacement from $\mathbf{X}$. In this study $B_{BS} = 999$
was used (the \texttt{mclust} default) to ensure robust variance estimation for each
of the illustrative data sets. It is computationally feasible to run a greater number
of bootstrap samples if required. See \cite{andrews2000three} for a formal guide to
choosing the number of bootstrap samples across a range of applications.
\end{itemize}

\item
\label{alg:zinitial}
For each sample $b = 1,2,\ldots,B$, construct the associated
initialization matrix of group membership probabilities
$\mathbf{\hat{Z}}_b$. This is populated with the values from the
$\mathbf{\hat{Z}}_{\tilde{M}}$ matrix formed using the full data that correspond to
each observation sampled. This circumvents the problem of label
switching (the problem that the likelihood is invariant under a permutation of the labels assigned to the mixture components) that would otherwise have to be explicitly undone at the end of the algorithm. It can be verified that this step successfully negates the possibility of label switching by checking the ordering of the sizes of the $\btau$ and $\bmu$ component probability and mean parameter estimates that emerge from each JK, BS, PB and WLBS sample fitted versus those of the optimal model. Across all data sets tested, this post processing step never failed to verify that label switching had been avoided through use of the  $\mathbf{\hat{Z}}_{\tilde{M}}$ matrix for initialization purposes and that the original component orderings remained unaltered.

\item
\label{alg:startingvals}
For each sample, calculate the MLEs of
$\tau_g$ and $\btheta_g$ under model $\tilde{M}$.
Initialization using the $\mathbf{\hat{Z}}_{\tilde{M}}$ matrix, as detailed in step $(\ref{alg:zinitial})$, greatly improves convergence times for each sample and the speed of the method as a whole. Empirical study showed that using random starts to initialize fit on resampled data sets makes minimal difference versus using the matrix $\mathbf{\hat{Z}}_{\tilde{M}}$ from the original fit, with convergent log likelihood values and parameter estimates in agreement under either approach. Use of the matrix $\mathbf{\hat{Z}}_{\tilde{M}}$ from the original model fit is merely preferred for purposes of computational efficiency and to circumvent the threat of label switching. It should be noted that, technically, starting the algorithm from the original fit is invalid, because it uses information that is not available when running the original Gaussian mixture maximum likelihood estimator on the data.

\item Estimate the (co)variance of any model parameter $\psi$:

\begin{itemize}
\item The jackknife estimate of a parameter's variance,
$\sigma_{JK}^2(\psi)$, is equal to the sample variance of the
$B_{JK}$ values of $\psi$ multiplied by the constant term
$\displaystyle\frac{(n-1)}{n}$, where $\overline{\psi}_{JK}$ denotes the jackknife sample mean. A move from the delete-$1$ jackknife to the general delete-d jackknife means that each sample formed contains fewer observations than in the delete-1 case: $(n-d)$ versus $(n-1)$. However, there is a larger number of samples available in the delete-$d$ case: $n \choose d$ as opposed to $n$. The net effect is that the delete-$d$ approach can produce superior estimates of variance for non-smooth statistics such as the median or quantiles. However, for estimating variance of smooth statistics such as the mean, covariance elements and proportions required in a model-based clustering context, the delete-$1$ variant is reliable and is markedly faster and more straightforward to implement \citep{shi1988note}:

\begin{equation}
\sigma_{JK}^2(\psi) =
\displaystyle\frac{(n-1)}{n}\displaystyle\sum_{m=1}^{B_{JK}}(\psi_m
- \overline{\psi}_{JK})^2.
\label{eq:JK_variance}
\end{equation}

\item The bootstrap estimate of a parameter's variance,
$\sigma_{BS}^2(\psi)$, is equal to the sample variance of the
$B_{BS}$ values of $\psi$ calculated across the bootstrap samples, where $\overline{\psi}_{BS}$ denotes the bootstrap sample mean:

\begin{equation}
\sigma_{BS}^2(\psi) =
\displaystyle\frac{1}{(B_{BS}-1)}\displaystyle\sum_{m=1}^{B_{BS}}(\psi_m
- \overline{\psi}_{BS})^2. \label{eq:BS variance}
\end{equation}
\end{itemize}

The bootstrap and jackknife estimates of covariance between
parameter estimates can be calculated using analogous formulae.

\end{enumerate}

Using \texttt{mclust} to fit the pre-specified model $\tilde{M}$ to each JK or BS sample, and using the full data model $\mathbf{\hat{Z}}_{\tilde{M}}$ matrix for initialization (as described in step ($\ref{alg:startingvals}$)), means the algorithm provides a quick and accurate way of estimating parameter (co)variances. It must be noted that the inference proposal and estimation of parameter standard errors is conditional on the method of model selection. If the user chooses the correct model in advance, there is no validity problem. However the converse situation where the model selection process uses the data, which arises commonly in statistical modeling, is not without peril - inference ignoring prior model selection is technically invalid \citep{Leeb2005}. Nonetheless, it represents the standard approach across the existing methods of variance estimation detailed in Section $\ref{sec:methods}$ and across the wider spectrum of statistical inference.

Bootstrapping can be asymptotically consistent but does not provide general finite-sample guarantees. However, it is a viable option for obtaining confidence limits in cases where a normal approximation of a parameter's distribution is not appropriate \citep{davison1997bootstrap}. On the other hand, by definition, the bootstrap density carries reduced inferential information about the underlying parameter since not all observations are represented in a typical sample and estimates of variability based upon the samples are less reliable \citep{Pawitan2000}. This is related to the fact that for nonparametric resampling the distribution of a parameter estimate is discrete, even though it may be approximating a continuous distribution, leading to ``fuzziness'' versus a parametric approach. However the support of the distribution tends to be fairly dense for samples of any reasonable size and hence the discrete approximation can often be viewed as relatively benign \citep{davison1997bootstrap}. A further impediment to asymptotic consistency of bootstrapping in this application is the fact that model selection is not performed on each bootstrap sample, but rather the full data optimal model is fitted across all samples \citep{AndrewGuggenberger2008}. However empirical testing suggests that this has limited impact because in most cases the full data model remains the optimal one in the samples formed. This is particularly true under the JK, BS and PB approaches. The fact that it has some impact in terms of not always being able to fit the model under the BS approach constitutes a further criticism of this method of variance estimation versus the JK, PB and WLBS methods.

\subsection{The parametric bootstrap}
\label{sec:PB}

The parametric bootstrap estimate of a parameter's variance, $\sigma_{PB}^2(\psi)$, is calculated in a similar manner to the bootstrap estimate of the parameter's variance. However, in the case of the parametric bootstrap, the $B_{PB}$ values of $\psi$ required to estimate its variance are not generated via resampling of the original data, but rather through simulation of $B_{PB}$ new data sets from the fitted model. For initialization purposes $\mathbf{\hat{Z}}_{\tilde{M}}$ is again used when generating each simulated data set, to improve convergence speed and prevent the issue of label switching. For each new simulated data set, the original fitted model is applied to the observations and the value of $\psi$ is determined for that data set. Then $\sigma_{PB}^2(\psi)$ is calculated as the sample variance of the $B_{PB}$ values of $\psi$. The \texttt{mclust} default is to set $B_{PB} = 999$. This could be increased, if necessary, to handle cases where some clusters are sparsely populated - ensuring a sufficient number of observations is simulated from such clusters to allow robust variance estimation for their distributional parameters (albeit at increased computational overhead). Clearly the accuracy and validity of the parametric bootstrap approach relies on the assumption that the fitted model provides a good approximation to the true mechanism that generated the original data set \citep{efron1982}, with its precision declining substantially as the model deviates from the correct model \citep{mita2012}. The parametric bootstrap can be used to simulate large numbers of observations from the specified model, ensuring that a sufficient number of observations from small clusters arise to permit variance estimation for their associated parameters.

\subsection{The weighted likelihood bootstrap}
\label{sec:WLBS}

For cases where one or more of the clusters in the data set contain relatively few observations, it is likely that such clusters will be under-represented in some of the BS and PB (and potentially JK) samples formed. In extreme cases such clusters may be completely unrepresented in some of the samples formed. Consequently the estimation of parameter standard errors corresponding to these clusters via such sampling based methods will either be highly unstable or not possible. The weighted likelihood bootstrap (WLBS) approach is proposed here as an effective remedy in such circumstances.

The weighted likelihood bootstrap \citep{newton94} originated as a way to simulate approximately from a posterior distribution. In the context of a sampling based approach to variance estimation, the WLBS differs from the JK and BS in that every observation in the data set  $\mathbf{X}$ is `present' in each WLBS sample formed. The degree to which each observation is present is measured by its associated `weight'. Each weight $w_i$ ($i = 1, \ldots, n$) is simulated. As in \cite{newton94}, the uniform Dirichlet distribution is employed for the purposes of simulating the weights here. The implication of using the uniform Dirichlet in this capacity is that the weights are effectively being simulated from an exponential distribution, with scaling provided by the mean of the exponential draws. Other weighting distributions for observations could alternatively be used, for example those based on the number of observations present in the cluster to which an observation belongs, but were found to yield inferior performance in terms of attributing sufficient weight to clusters with few observations to allow stable and robust parameter estimation versus the full data model.

The shift to the weighted likelihood bootstrap approach requires that when fitting the model $\tilde{M}$ from the full data to the WLBS sampled data, a weighted form of the complete data likelihood ($\ref{eq:complete data
likelihood}$) is now maximized:

\begin{equation}
L_{wc} = \displaystyle\prod_{i=1}^n\displaystyle\prod_{g=1}^G\left[\tau_g f(x_i|\mu_g, \Sigma_g)\right]^{z_{ig} w_i}
\label{eq:complete weighted data likelihood}
\end{equation}

As with the original bootstrap method, outlined in Section $\ref{sec:jkbs}$, $999$ weighted likelihood bootstrap samples are formed for each of the
illustrative data sets by sampling $999$ weight vectors $w = (w_1, \ldots, w_n)$; this ensures robust variance estimation. Each sample formed again contains $n$ observations, but sampling with replacement is no longer employed -- all $n$ original observations are present in each of the $999$ samples formed, but each observation has associated weight $w_i$. Hence, the WLBS resolves the under-representation of small clusters that arises in the BS (and potentially JK) cases, as each WLBS sample includes all observations.

Variance estimates of model parameters are calculated under the WLBS approach in the same manner as in the BS method outlined in the algorithm in Section $\ref{sec:jkbs}$ -- with the exception that at step $(\ref{alg:sample})$ $B_{WLBS} = 999$ samples are formed by sampling weight vectors from the uniform Dirichlet distribution. Thus, the WLBS also provides a quick and accurate way of estimating parameter variances, even in the presence of small/overlapping clusters. As such, the WLBS approach provides a robust nonparametric alternative to the parametric bootstrap approach.

\vskip14pt
\section{Illustrative Data Sets}
\label{sec:data}

The application and performance of the JK, BS, PB and WLBS approaches to variance estimation detailed in Section $\ref{sec:methods}$ are demonstrated through three simulation studies and through the use of two well established clustering data sets, the \emph{Old Faithful} data and the \emph{Thyroid} data.

\subsection{Simulated data sets}
\label{sec:simdata}

Three simulation settings are used to illustrate the proposed sampling based approaches to variance estimation, and to assess and compare their performance and computational efficiency.

\subsubsection{Simulation Setting One and Simulation Setting Two.}
\label{sec:simdata12}

\begin{figure}
\begin{center}
\begin{tabular}{cc}
\subfigure[]{\label{fig:Fig1a}\includegraphics[width=6cm,height=7cm,angle=270]{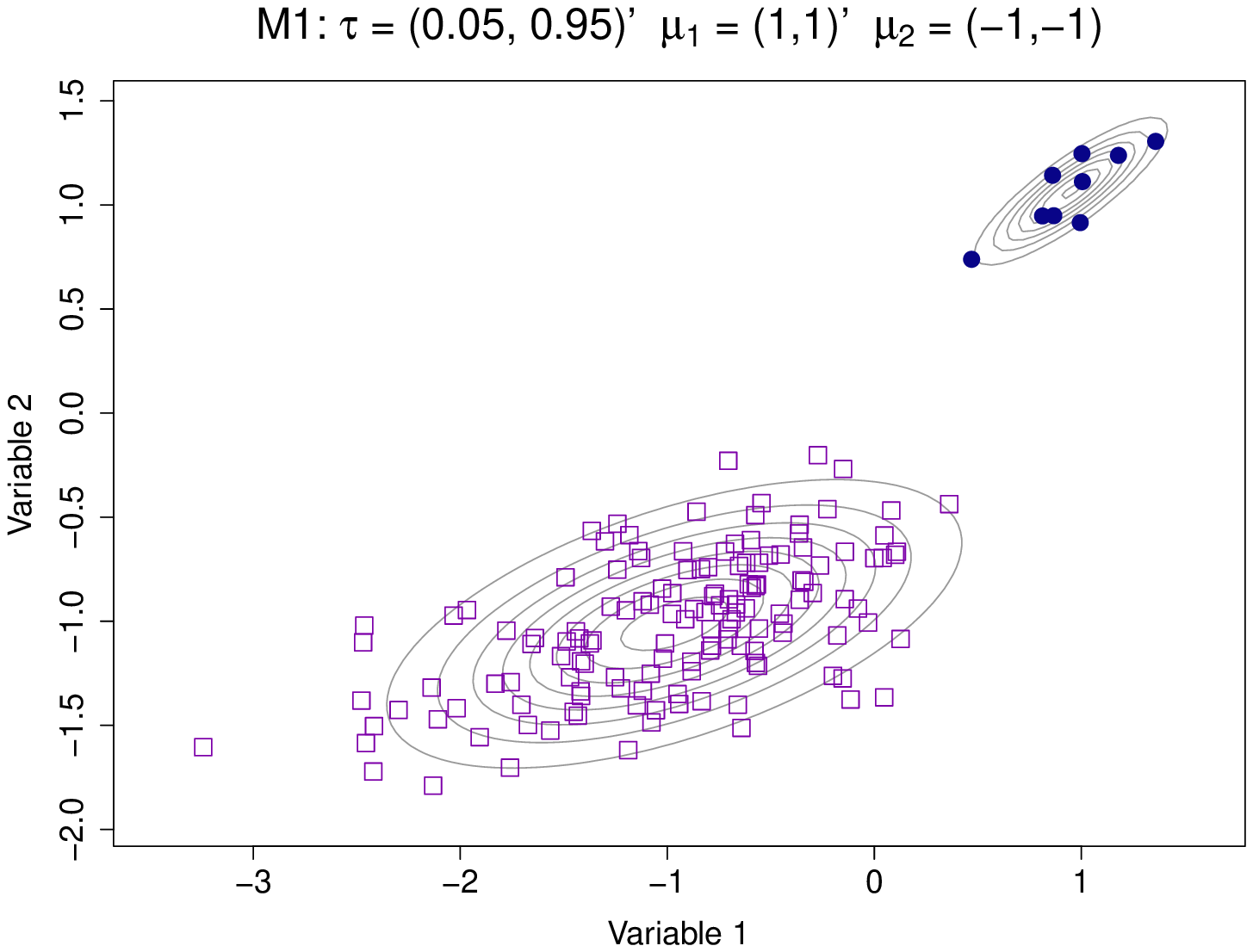}}&
\subfigure[]{\label{fig:Fig1b}\includegraphics[width=6cm,height=7cm,angle=270]{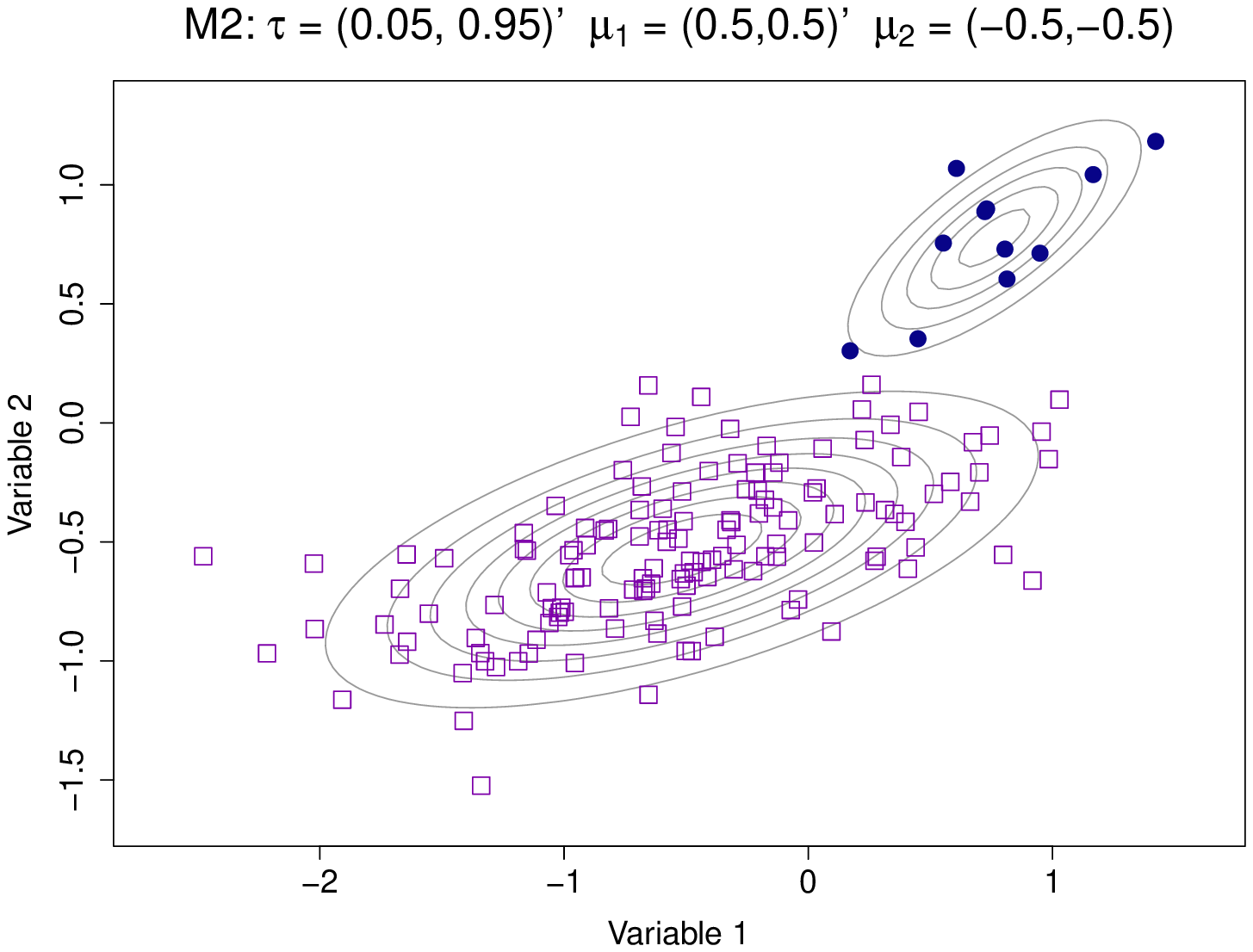}}\\
\subfigure[]{\label{fig:Fig1c}\includegraphics[width=6cm,height=7cm,angle=270]{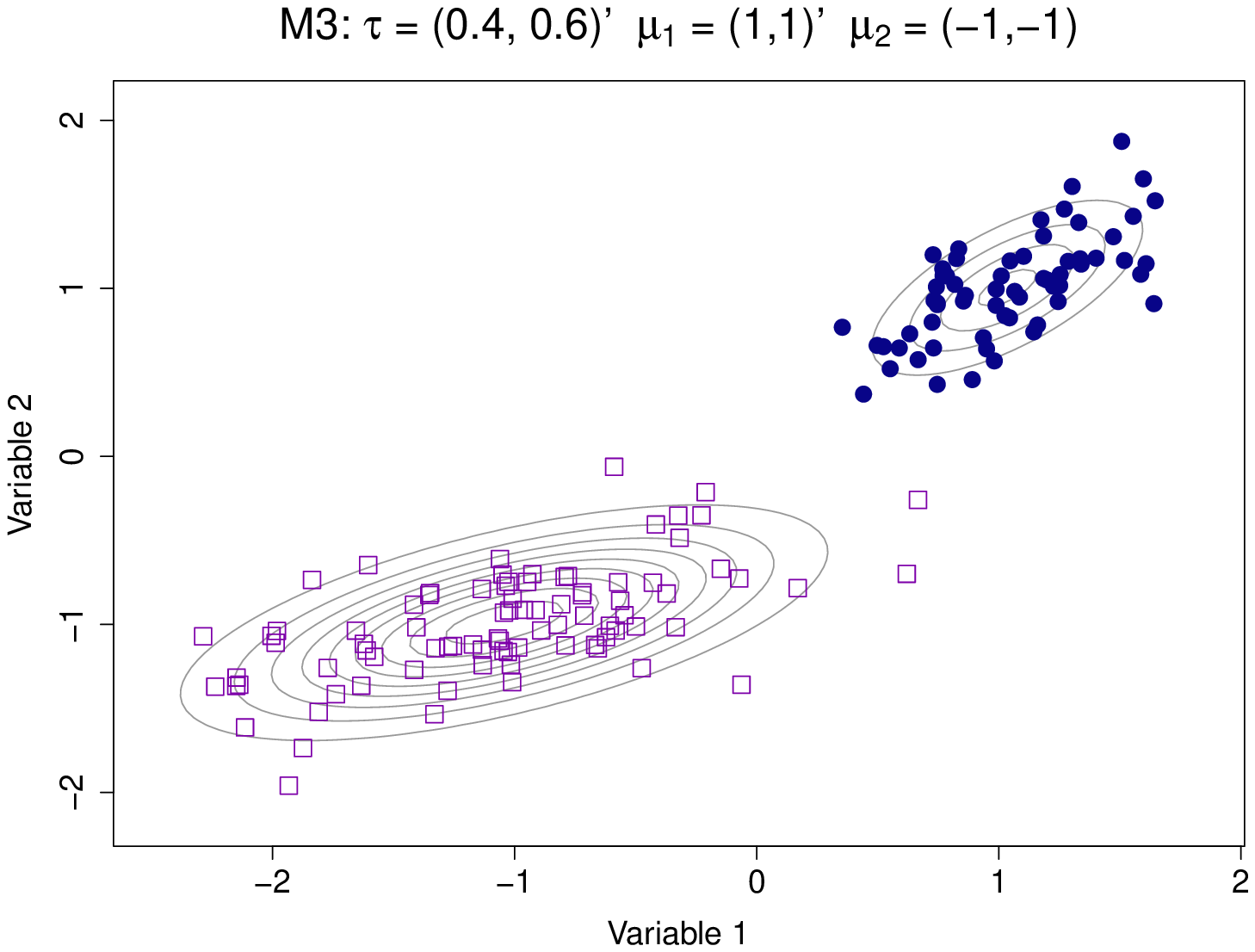}}&
\subfigure[]{\label{fig:Fig1d}\includegraphics[width=6cm,height=7cm,angle=270]{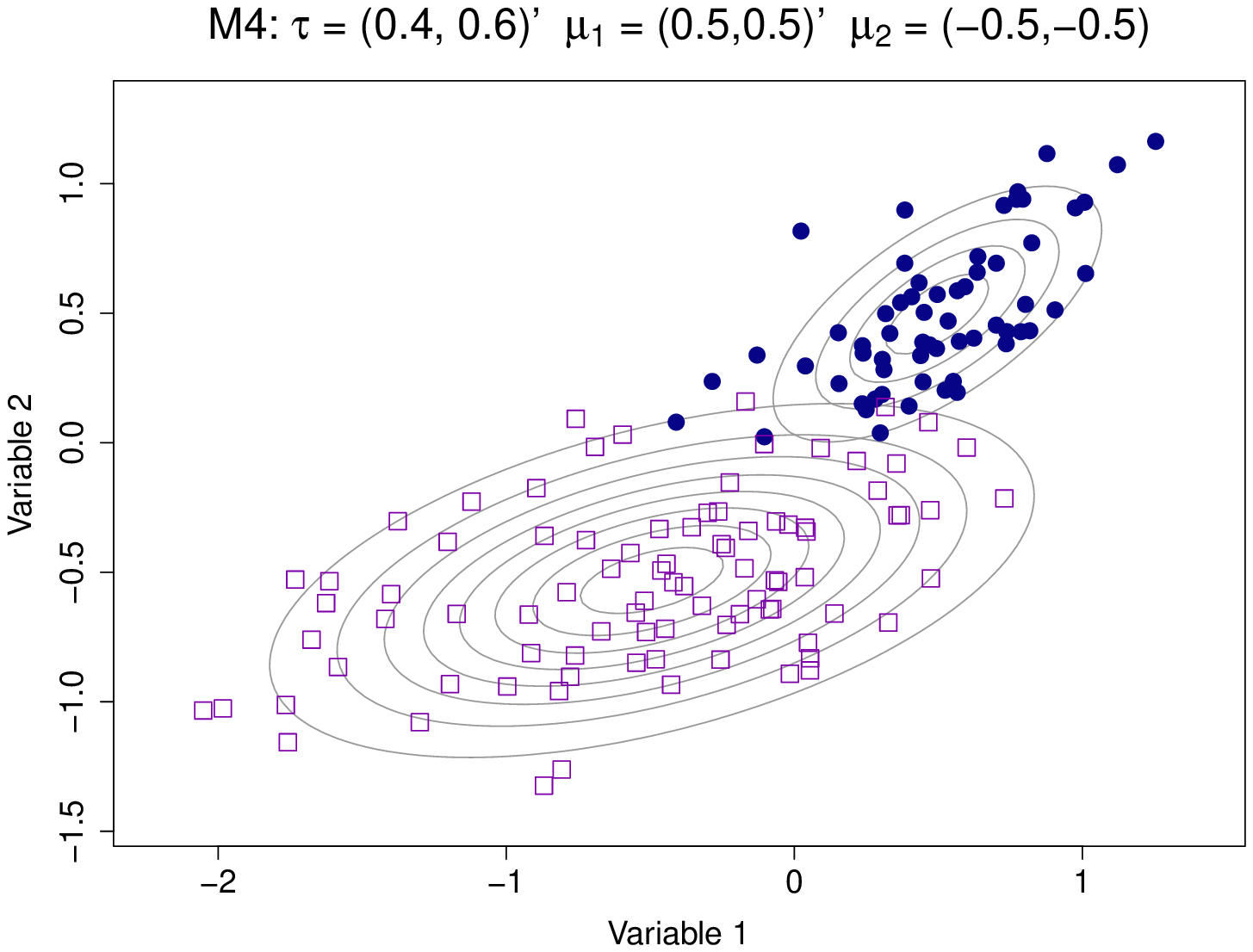}}
\end{tabular}
\caption{Scatter plot for a single simulated data set ($n = 150$) from each of the four models ((a) M1, (b) M2, (c) M3 and (d) M4) considered in the $G = 2$ simulation setting. M1 and M2 consider the case where a small cluster is present; M3 and M4 consider more equally sized clusters. M1 and M3 consider non-overlapping clusters whereas M2 and M4 consider overlapping clusters. True parameter settings are detailed above each figure.}
\label{fig:simulated data G2}
\end{center}
\end{figure}

\begin{figure}[!h]
\begin{center}
\begin{tabular}{cc}
\subfigure[]{\label{fig:Fig2a}\includegraphics[width=6cm,height=7cm, angle=270]{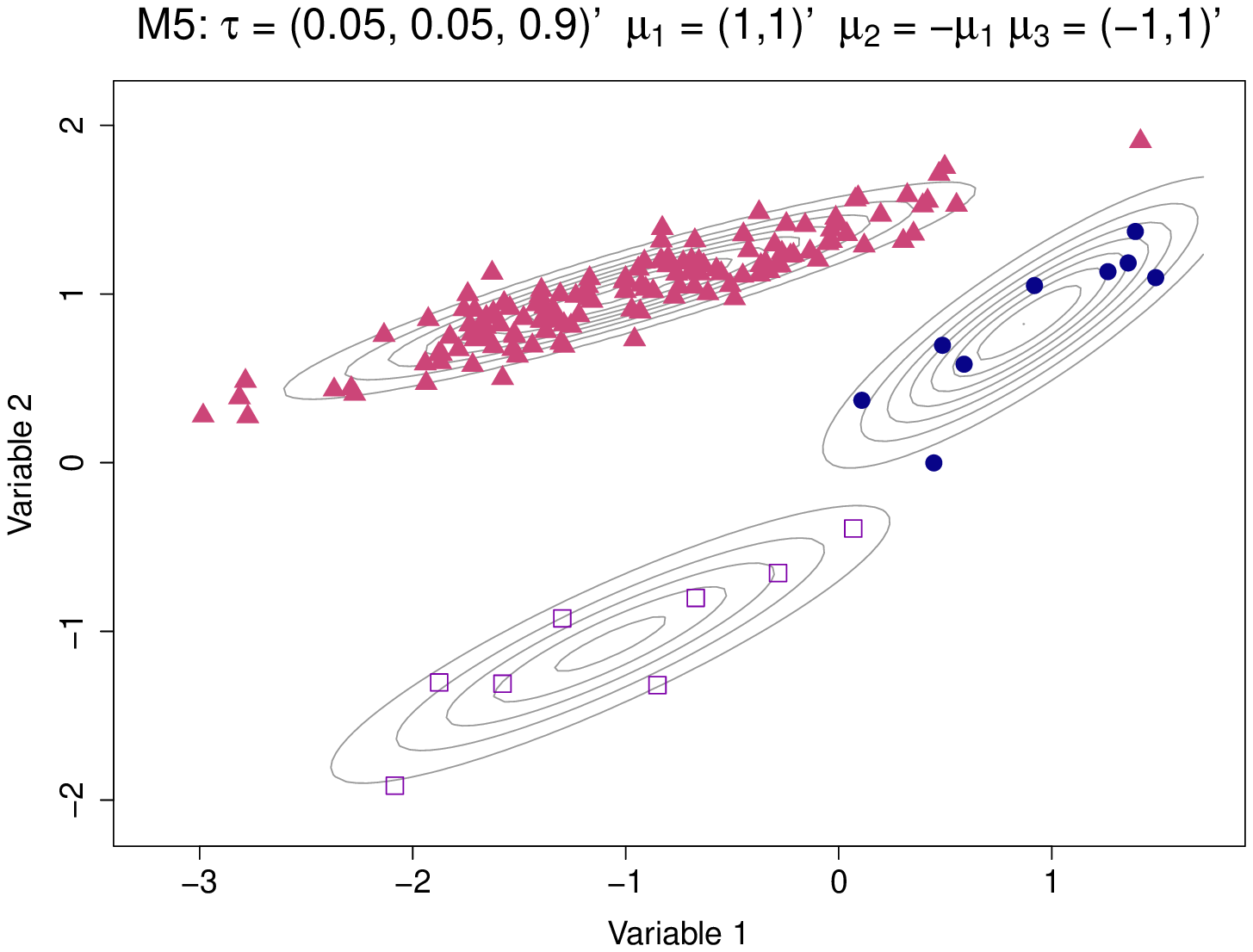}}&
\subfigure[]{\label{fig:Fig2b}\includegraphics[width=6cm,height=7cm, angle=270]{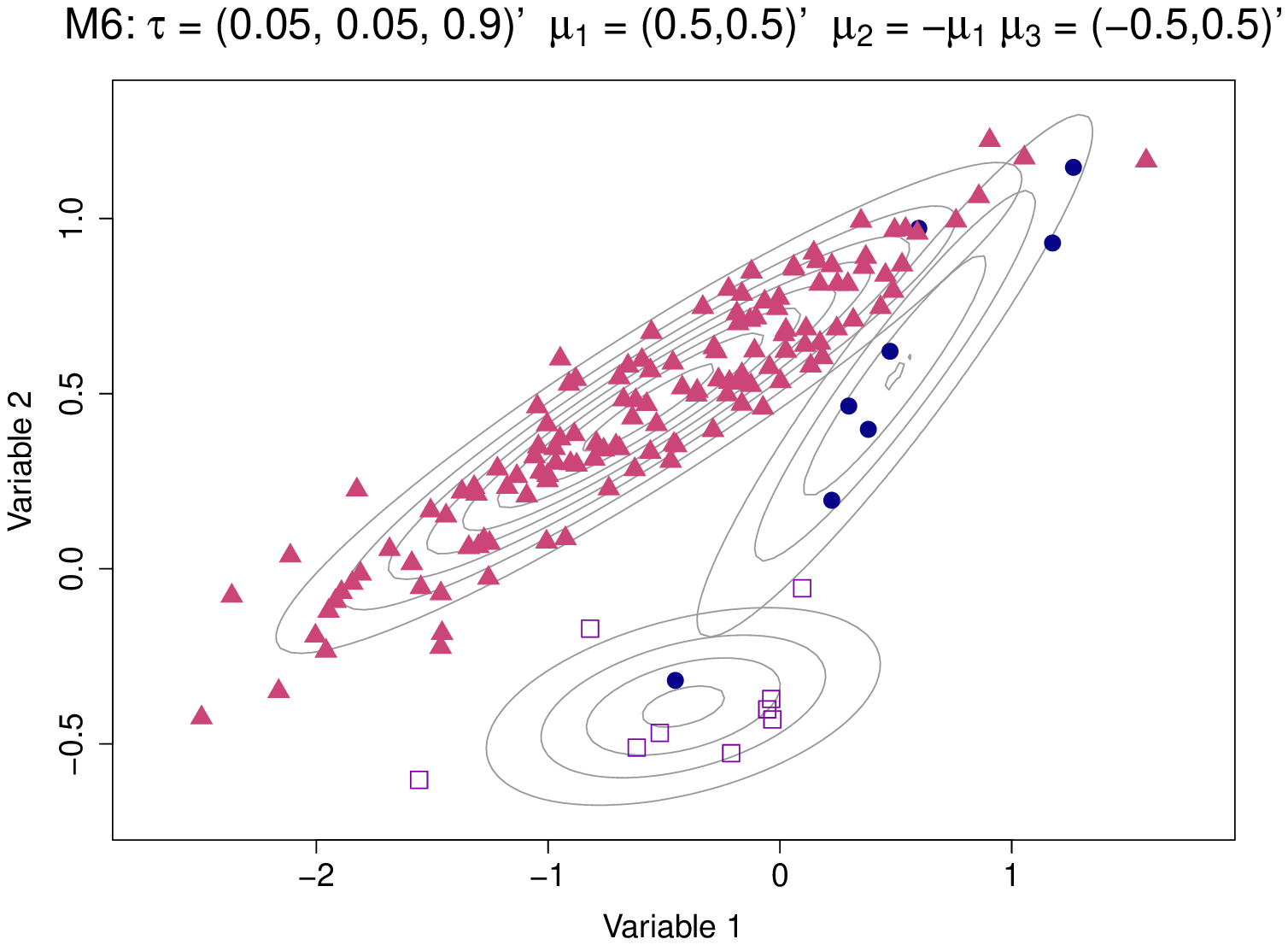}}\\
\subfigure[]{\label{fig:Fig2c}\includegraphics[width=6cm,height=7cm, angle=270]{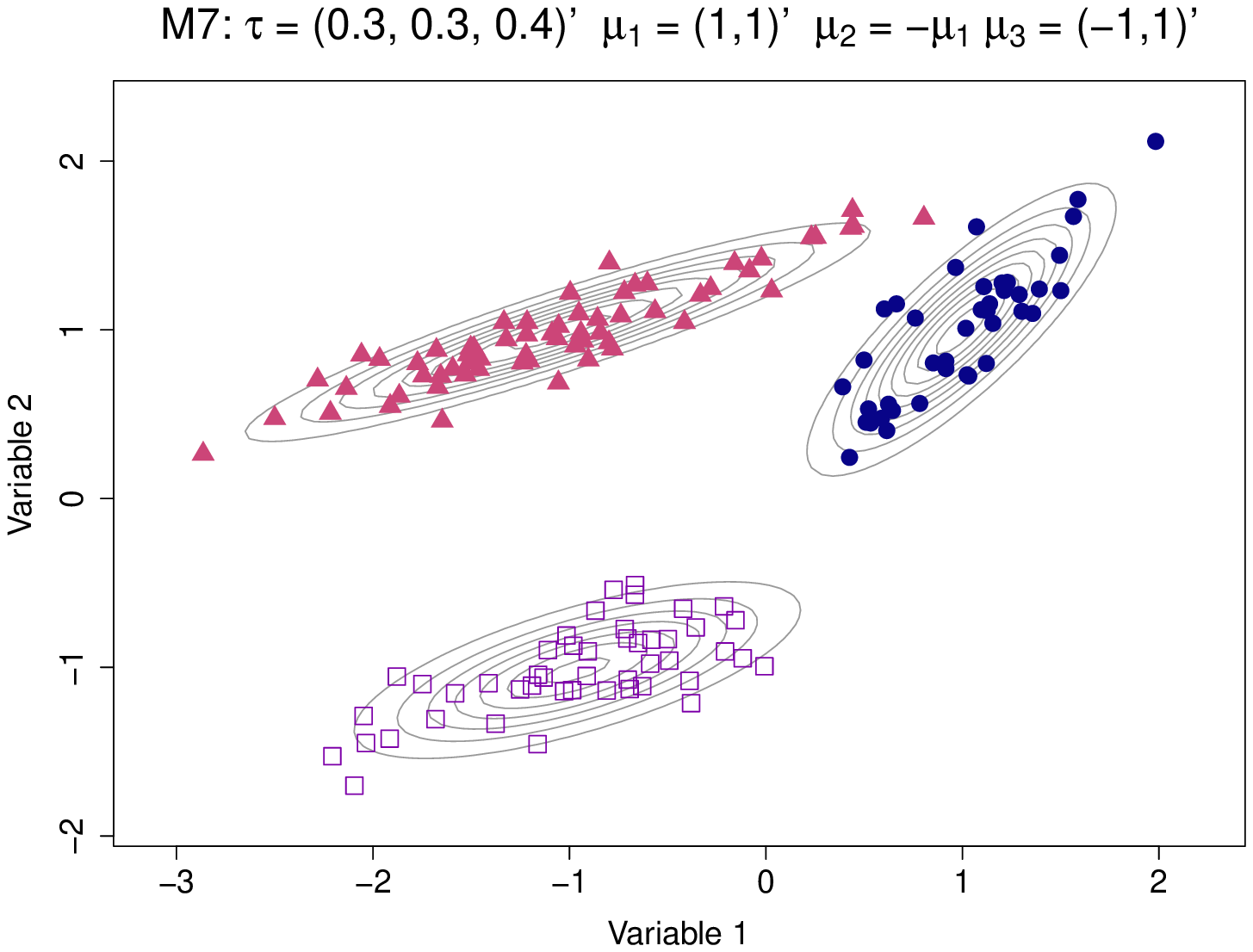}}&
\subfigure[]{\label{fig:Fig2d}\includegraphics[width=6cm,height=7cm, angle=270]{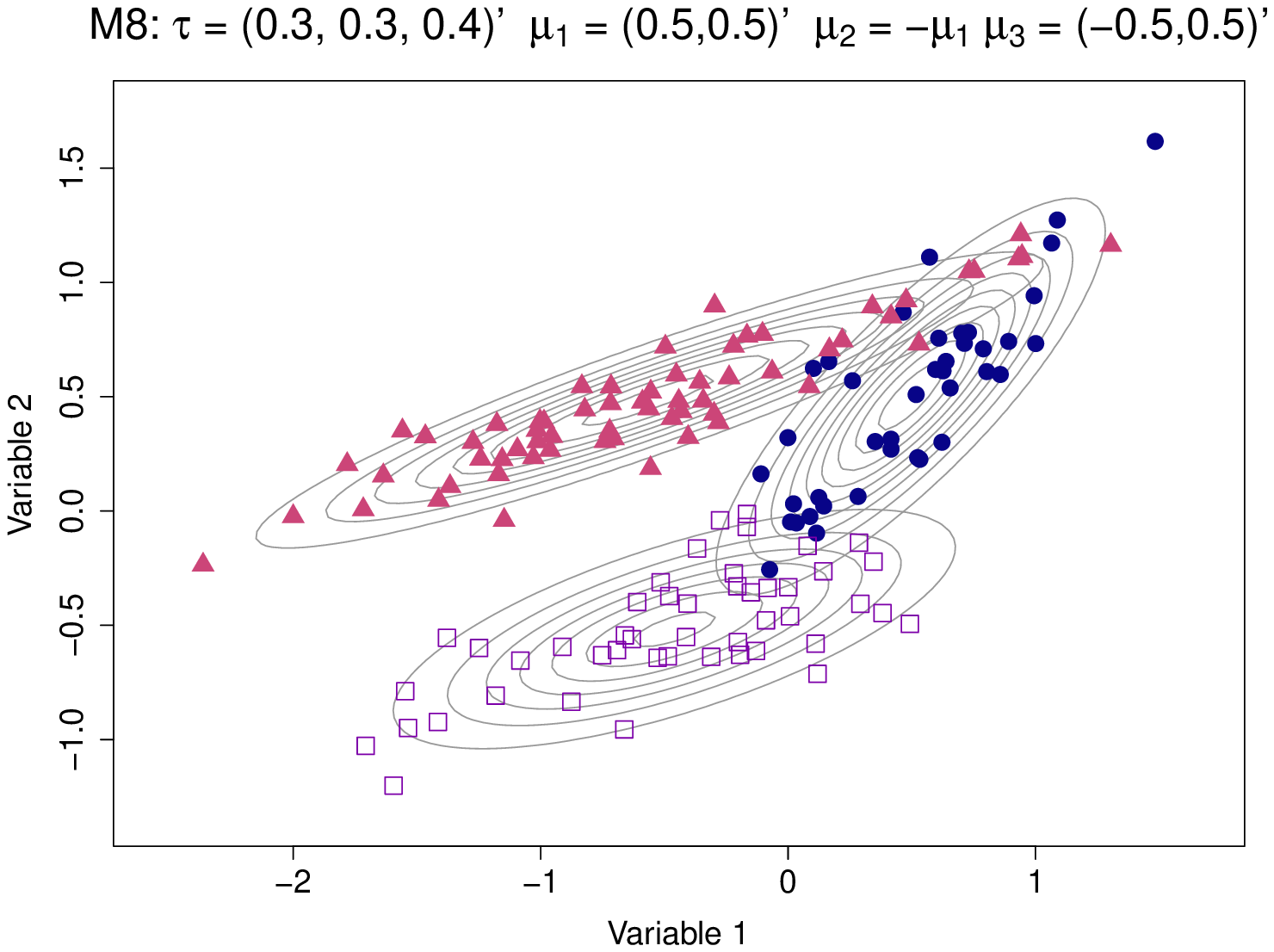}}
\end{tabular}
\caption{Scatter plot for a single simulated data set ($n = 150$) from each of the four models ((a) M5, (b) M6, (c) M7 and (d) M8) considered in the $G = 3$ simulation setting. M5 and M6 consider the case where small clusters are present; M7 and M8 consider more equally sized clusters. M5 and M7 consider non-overlapping clusters whereas M6 and M8 consider overlapping clusters. True parameter settings are detailed above each figure.}
\label{fig:simulated data G3}
\end{center}
\end{figure}

Two illustrative simulation settings are considered here to clearly expose the proposed sampling based approaches to variance estimation, and to assess and compare their performance. Both simulation settings consider a mixture of Gaussians model, one in which $G = 2$ and one in which $G = 3$. In both settings, for illustrative purposes, the number of variables $p = 2$, and in order to thoroughly test performance a small sample size of $n = 150$ was used. Within each simulation setting, four different models are considered, as illustrated in Figures $\ref{fig:simulated data G2}$ and $\ref{fig:simulated data G3}$. In brief, the four models examined in each setting consider differently sized clusters with different degrees of cluster separation. The covariance structure used varies between clusters in all instances (i.e. the `VVV' \texttt{mclust} model is used).

The true cluster covariance matrices, $\Sigma_1^{TRUE}$ and $\Sigma_2^{TRUE}$, for $G = 2$ in \emph{Simulation Setting One}, for models M1, M2, M3 and M4 are:

\begin{center}
\begin{tabular}{llll}
$\Sigma_1^{TRUE}$ &
$\hspace{-2mm}= \left(\begin{array}{cc}
0.12 &       0.09 \\
0.09   &     0.12\end{array}\right)$
 &
$\,\,\,\,\,\,\,\,\,\,\Sigma_2^{TRUE}$ &
$\hspace{-2mm}= \left(\begin{array}{cccc}
0.47 &       0.13   \\
0.13   &     0.11   \end{array}\right)$
\end{tabular}
\end{center}

The true cluster covariance matrices, $\Sigma_1^{TRUE}$, $\Sigma_2^{TRUE}$ and $\Sigma_3^{TRUE}$, for $G = 3$ in \emph{Simulation Setting Two}, for models M5, M6, M7 and M8 are:

\begin{center}
\begin{tabular}{llllll}
$\Sigma_1^{TRUE}$ &
$\hspace{-2mm}= \left(\begin{array}{cc}
0.12 &       0.09 \\
0.09   &     0.12\end{array}\right)$
 &
$\,\,\,\,\,\,\,\,\,\,\Sigma_2^{TRUE}$ &
$\hspace{-2mm}= \left(\begin{array}{cccc}
0.39 &       0.15   \\
0.15   &     0.10   \end{array}\right)$
 &
$\,\,\,\,\,\,\,\,\,\,\Sigma_3^{TRUE}$ &
$\hspace{-2mm}= \left(\begin{array}{cccc}
0.53 &       0.20   \\
0.20   &     0.09   \end{array}\right)$
\end{tabular}
\end{center}

\subsubsection{Simulation Setting Three}
\label{sec:simdata3}

For illustrative and reporting clarity the simulation settings described in Section \ref{sec:simdata12} are low dimensional in nature. A further simulation study is also conducted which involves higher dimensional scenarios. The purpose of this additional simulation study is to further explore the performance and computational features of the JK, BS, PB and WLBS approaches to parameter variance estimation, in more complex scenarios.

A mixture of Gaussians model in which the number of clusters $G = 5$ is considered, where the cluster probabilities are set to be $\tau = (0.07, 0.07, 0.22, 0.27, 0.37)'$. Two settings for the number of observations $n$ are considered ($n = 500$ and $n = 700$). The number of variables $p$ considered is high within the context of dimensionality that the popular \textbf{R} package \texttt{mclust} \citep{R2017, fraley02, fraley12} can reasonably handle in terms of computational speed. Here three settings ($p = 15$, $p = 20$ and $p = 25$) are considered. Within each setting the covariance structure used varies between clusters in all instances (i.e. the `VVV' \texttt{mclust} model is used), and some small clusters are present. There is also overlap between the clusters; Figures $\ref{fig:M91}$, $\ref{fig:M92}$ and $\ref{fig:M93}$ in Appendix \ref{app:sim3plots} illustrate this to some degree through pairs plots from a single simulated data set for which $n = 500$, $p = 25$ and $G = 5$.

\subsection{The \emph{Old Faithful} data}
\label{sec:oldfaithfuldata}

\begin{figure}[!h]
\begin{center}
\includegraphics[width = 110mm, angle = 270]{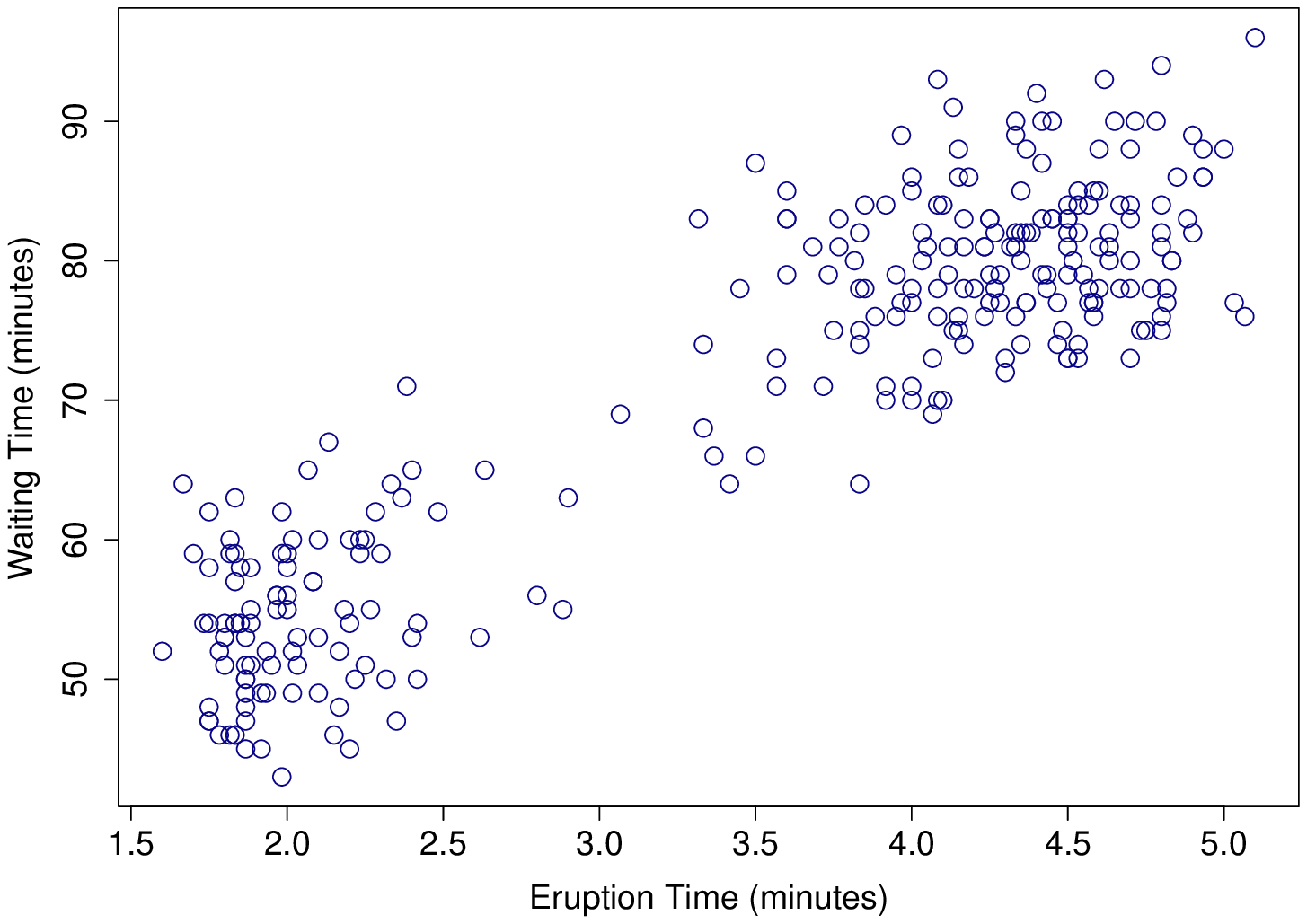}
\caption{Scatter plot of the \emph{Old Faithful} data.}
\label{fig:Faithful data}
\end{center}
\end{figure}

The frequently utilised \emph{Old Faithful} data set is comprised of bivariate observations for $272$ eruptions
of the \emph{Old Faithful} geyser in Yellowstone National Park
\citep{azzalini1990}. Each observation records the eruption duration
and the waiting duration until the next eruption, both measured in
minutes; the data are illustrated in Figure $\ref{fig:Faithful data}$. This is a classic
test case for any clustering methodology because the data are
multimodal. However, there are no `true' group labels available -- the presence of various numbers of groups has been suggested, depending on the clustering rule applied.

\subsection{The \emph{Thyroid} data}
\label{sec:thyroiddata}

\begin{figure}[!h]
\begin{center}
\includegraphics[width = 110mm, angle = 270]{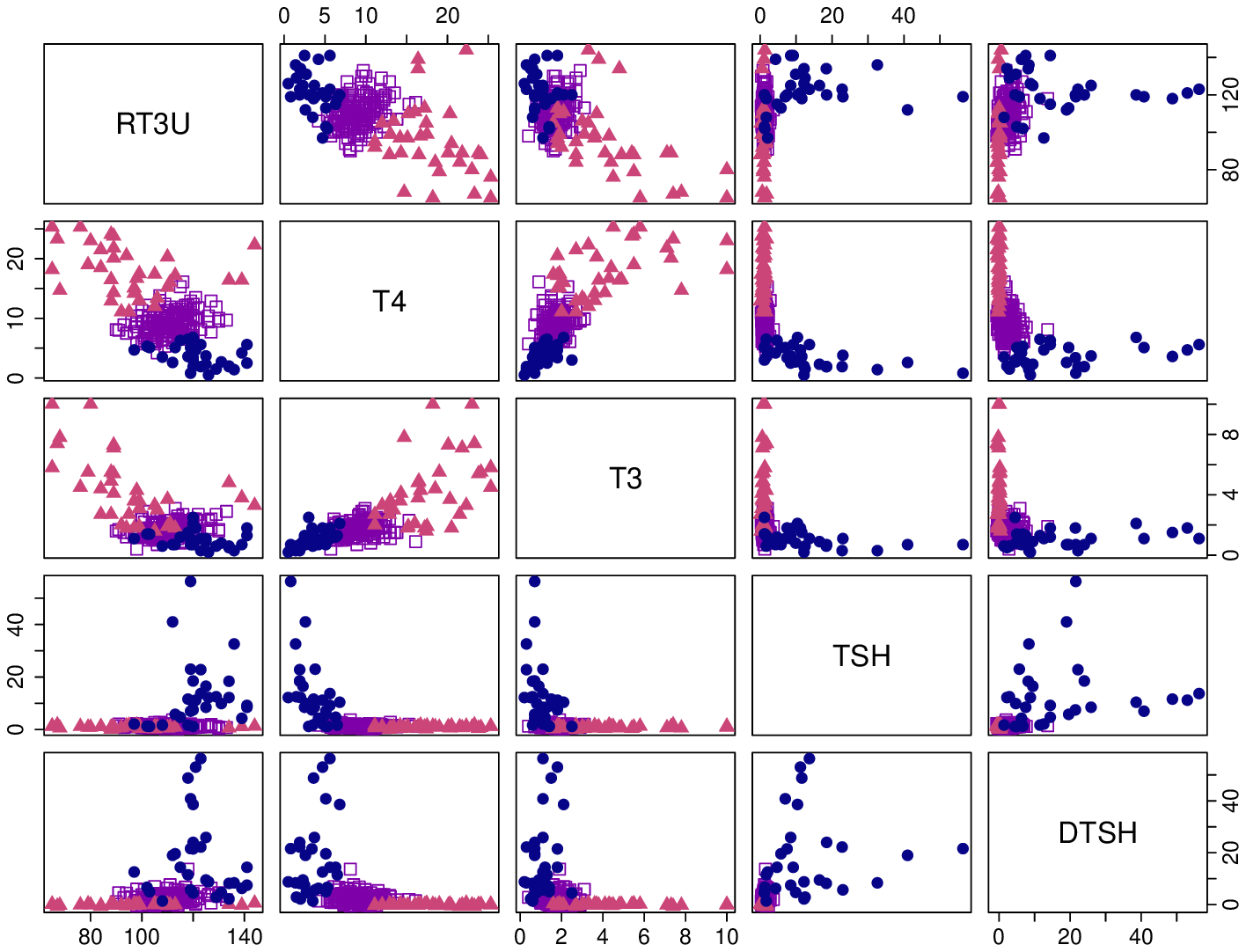}
\caption{Pairs plots of the \emph{Thyroid} data across the five laboratory tests administered. Observations in navy denote the ``normal'' diagnosis of eurothyroidism, while those in purple and pink denote diagnoses of hypothyroidism and hyperthyroidism respectively.}
\label{fig:Thyroid data}
\end{center}
\end{figure}

The \emph{Thyroid} data set is comprised of data on five laboratory tests administered to a sample of $215$ patients. The tests are: percentage
T3 resin uptake test (RT3U); total serum thyroxin as measured by the isotopic displacement method (T4); total serum triiodothyronine as measured by radioimmuno assay (T3); basal thyroid-stimulating hormone as measured by radioimmuno assay (TSH); maximal absolute difference of TSH value after injection of $200$ micrograms of thyrotropin-releasing hormone as compared to the basal value (DTSH). The tests are used to predict whether a patient's thyroid can be classified as euthyroidism (normal thyroid gland function), hypothyroidism (underactive thyroid not producing enough thyroid hormone) or hyperthyroidism (overactive thyroid producing and secreting excessive amounts of the free thyroid hormones T3 and/or thyroxine T4). Diagnosis of thyroid operation was based on a complete medical record, including anamnesis, scans and other methods and is included in the data set. The data are illustrated in Figure $\ref{fig:Thyroid data}$. Observations in navy denote the ``normal'' diagnosis of eurothyroidism, while those in purple and pink denote diagnoses of hypothyroidism and hyperthyroidism respectively. See \cite{coomans1983comparison} for further details.

\section{Results}
\label{sec:results}

The application and performance of the JK, BS, PB and WLBS approaches to variance estimation are illustrated using the data sets outlined in Section $\ref{sec:data}$. For both the simulation study and the \emph{Old Faithful} and \emph{Thyroid} data sets parameter estimates are presented, as are their associated standard errors computed under the JK, BS, PB and WLBS methods using a mixture model where each component is multivariate Gaussian.

\subsection{Simulation study}
\label{sec:sim study results}

The simulation settings described in Section $\ref{sec:simdata}$ serve as a means of assessing the accuracy of the sampling-based approaches to parameter variance estimation, in different clustering scenarios.

\subsubsection{Simulation Setting One: $G = 2$.}
\label{sec:case one sim study results}

For the first simulation setting, where two clusters are present, Table $\ref{tab:sim case one tau results}$ presents the cluster probability estimates and associated standard errors, based on a single simulated data set under each of the models M1, M2, M3 and M4 with $n = 150$, for the JK, BS, PB and WLBS methods. These cluster probability estimates are the mean values of $\tau$ for each group across all resampled versions of the data constructed, for each of the four resampling methods considered. It is clear that, for models M1, M3 and M4, all methods produce cluster probability estimates that are very close to the true values and that the standard errors are relatively small. Results are poor however for the more challenging model M2, in which a small cluster is present and the clusters overlap. All approaches over-estimate the number of observations belonging to the smaller cluster in this case, and the standard errors are larger than in model M1, which also has a small cluster present. Also notable is that, when moving from M3 to M4,
while parameter estimates appear unchanged, the standard errors increase due to the increased cluster overlap and the resulting poor clustering performance (in terms of misclassification rate). Similar observations arose when examining the cluster mean and covariance estimates and standard errors for the four models.

Under the BS and PB, a total of $B_{BS} = B_{PB} = 999$ samples was requested in this study. Here, under models M1 and M2, a number of additional samples was required in order to achieve this total, attributable to the inability to fit the optimal model $\tilde{M}$ to some of the sampled data sets. Inability to fit a model is caused by non-convergence of the EM algorithm resulting from reaching a set of parameter estimates where the covariance matrix cannot be inverted. This is attributable to the random nature of the data selected to form the affected samples. It particularly affects the BS approach since, by design, many of the original observations may not be included in any given sample formed. Resampling may produce samples that have no or few observations from the small cluster present in the M1 and M2 model settings. Likewise, a BS sample may include the same observation multiple times. Finally, the optimal model may not be a good fit to the underlying data generating mechanism and thus fitting it to a sampled data set may not be possible (which can greatly compromise the PB approach). Any one of these issues, or a combination of them, can lead to the computational instability described. Empirical study shows that this issue cannot be overcome using a different initialization method, such as random initializations, for the EM algorithm, since the issue is inherent to the data selected. Such non-fitting issues can potentially occur in the JK and WLBS approaches (but empirically do so much less frequently in the JK case, and never in the WLBS case).

Hence it is appropriate to also consider and report the \emph{effective number of samples} ($EB$) drawn. In the BS, PB and WLBS cases, $EB$ is defined to be the number of sampled data sets constructed in order to compute the variance estimates using the requested $B_{BS}$, $B_{PB}$ and $B_{WLBS}$) number of samples. To avoid entering an infinite loop of drawing non-estimable bootstrap samples an upper limit of $B_{BS} \times 10 = 9990$ is set (and likewise for the PB and WLBS); once this number of non-estimable samples has been drawn the algorithm is terminated and reported variance estimates are based on the successfully estimated models only.
In the JK case, drawing more than $n$ samples is clearly not feasible and so non-estimable samples are discarded. Thus in the JK approach $EB$ denotes the actual number of sampled data sets involved in the final computation of the reported variance estimates. Table $\ref{tab:sim case one tau results}$ shows that the BS and PB require an EB slightly above the default $999$ data sets for models M1 and M2. No computational issues requiring additional samples to be drawn are encountered for any of the other models and variance estimation methods considered.

\begin{table}[h]
\caption{Cluster probability estimates
(with associated standard errors) for a data set simulated under each of the four models detailed in Figure $\ref{fig:simulated data G2}$ where there are $G = 2$ multivariate Gaussian components, under each of the variance estimation approaches. $^\dagger$ For M1, $EB_{BS} = 1001$ and $EB_{PB} = 1002$. $^\ddagger$ For M2, $EB_{BS} = 1005$ and $EB_{PB} = 1000$. The true values are $0.05$ and $0.95$ for M1 and M2 and $0.4$ and $0.6$ for M3 and M4.}
\centering
\label{tab:sim case one tau results}
\begin{tabular}{lllll}
\\
  & \multicolumn{1}{c}{$ \mathbf{\tau_{JK}} $} & \multicolumn{1}{c}{$ \mathbf{\tau_{BS}} $} & \multicolumn{1}{c}{$\mathbf{\tau_{PB}} $} & \multicolumn{1}{c}{$\mathbf{ \tau_{WLBS}} $}  \\
\bf{M1} & \{0.07 (0.02), 0.93 (0.02)\} & \{0.07 (0.02), 0.93 (0.02)\}$^\dagger$ & \{0.07 (0.02), 0.93 (0.02)\}$^\dagger$ & \{0.08 (0.03), 0.92 (0.03)\}\\
\bf{M2} & \{0.13 (0.04), 0.87 (0.04)\} & \{0.13 (0.05), 0.87 (0.05)\}$^\ddagger$ & \{0.14 (0.05), 0.86 (0.05)\}$^\ddagger$ &\{0.14 (0.05), 0.86 (0.05)\} \\
\bf{M3} & \{0.38 (0.04), 0.62 (0.04)\} & \{0.38 (0.04), 0.62 (0.04)\} & \{0.38 (0.04), 0.62 (0.04)\}  & \{0.38 (0.04), 0.62 (0.04)\} \\
\bf{M4} & \{0.38 (0.06), 0.62 (0.06)\} & \{0.38 (0.08), 0.62 (0.08)\} & \{0.37 (0.06), 0.63 (0.06)\} & \{0.38 (0.09), 0.62 (0.09)\} \\
\end{tabular}
\end{table}

To further assess the performance of the methods, $1000$ data sets were generated under each of the models M1, M2, M3 and M4. For each data set an approximate $95$\% confidence interval (mean $\pm 2$ standard errors for the BS, PB and WLBS approaches and using pseudo-values for the JK approach) was formed for each model parameter. The number of data sets for which the confidence interval contained the true parameter value was recorded; these coverage results are presented in Table $\ref{tab:tau1_sim_study_table 1}$ for the first group membership probability parameter $\tau_1$.

Coverage results are good across all the sampling based approaches under models M1, M3 and M4. This is not the case for M2 where coverage declines markedly for all four resampling methods, due to the presence of a small, overlapping cluster. The JK fares best for M2, followed by the WLBS. The BS and PB both perform poorly. Similar trends were observed when examining the coverage results for the other model parameters.

The BS and PB again have fitting difficulties with models M1 and M2 because resampling and simulation respectively produce samples to which it was not possible to fit the optimal model $\tilde{M}$ in some instances. However, good coverage results are obtained for M1 (but not M2) after drawing additional samples until the required $B_{BS} = B_{PB} = 999$ fits are achieved. To a much lesser extent the JK also has difficulties with models M1 and M2 where it is again not possible at all times to fit the required model to the sampled data set; drawing additional samples is clearly not possible in the JK setting. While the WLBS appears to perform consistently well in terms of fitting ability, it should be noted that even if the BS approach needs to draw additional samples due to non-fitting issues, in low dimensional settings the BS is typically computationally more efficient than the WLBS approach because the latter requires the computation of the log-likelihood as a weighted sum of densities for each data point.

\begin{table}[!h]
\caption{Coverage results (i.e. the proportion of data sets for which the true probability of membership of cluster $1$ is within the JK, BS, PB and WLBS $95$\% confidence intervals for the models M1, M2, M3 and M4) where there are $G = 2$ multivariate Gaussian components. The first column (`$\tilde{M}$ Fits') reports the number of simulated data sets to which it was possible to fit the optimal model $\tilde{M}$; a total of $1000$ data sets were simulated under each model setting. The `Non-fit' columns detail the average number of resamples that did not converge within each sampling procedure, with standard deviations of the counts given in parentheses. Note that in this study $B_{JK} = n = 150$ and $B_{BS} = B_{PB} = B_{WLBS} = 999$.} 
\vspace{0.5cm}
\label{tab:tau1_sim_study_table 1}
\centering
\begin{tabular}{l|c|lc|lc|lc|lc|}
& $\tilde{M}$ & \multicolumn{2}{c|}{\bf{JK}} & \multicolumn{2}{c|}{\bf{BS}}& \multicolumn{2}{c|}{\bf{PB}}&\multicolumn{2}{c|}{\bf{WLBS}}\\
& Fits & Non-fit & Coverage & Non-fit & Coverage & Non-fit & Coverage & Non-fit & Coverage\\\hline
\bf{M1} & 977 & 0.04 (0.31) & 0.969 & 130.93 (369.86) & 0.957 & 41.02 (90.07) & 0.949 & 0.00 (0.00) & 0.948 \\
\bf{M2} & 1000 & 0.02 (0.21)&0.748 &  46.20 (208.55)&  0.508 & 9.73 (45.77)  & 0.488 & 0.00 (0.00)  & 0.601\\
\bf{M3} & 1000 & 0.00 (0.00) & 0.953 &   0.00 (0.00) & 0.945 &  0.00 (0.00) & 0.948  &  0.00 (0.00) & 0.947 \\
\bf{M4} &  1000 & 0.00 (0.00) & 0.935 &  0.01   (0.13) & 0.980 & 0.00 (0.00) &  0.968 & 0.00 (0.00) &  0.983\\
\end{tabular}
\end{table}

The accuracy of the standard errors obtained under the sampling based approaches can be assessed by a comparison to the true analytically derived standard errors when they are available. The missing information principle (MIP) \citep{tanner_2012} is used to analytically derive the true standard errors in a tractable and illustrative one dimensional setting. A Newton-based numerical method (NM) is also employed to compute standard errors as an additional comparison. Furthermore, numerical derivatives and Hessians are available via the popular \textbf{flexmix} package \citep{grun2007fitting}, as supported by the theory of \cite{hong2015extremum}. The challenging simulated data setting of model M2, variable $2$ is chosen for these comparisons -- one of the two underlying clusters is small and the clusters overlap in this variable. Table \ref{tab:analyticSEs} reports the standard errors computed under the JK, BS, PB, WLBS, MIP, NM and \textbf{flexmix} approaches. The sampling based approaches perform well, in that the standard errors they return are very close to those obtained analytically and numerically. The standard errors returned by all four sampling based approaches are very close to the truth for the larger cluster $2$ (i.e. for parameters $\mu_2$ and $\sigma_2$); in the case of the small cluster $1$, the WLBS and PB approaches sometimes report smaller standard errors than the truth, whereas the JK and BS approaches return slightly inflated standard errors throughout. The performance of the \textbf{flexmix} approach is similar to that of the WLBS and PB, indicating its suitability as a good competitor to the sampling based approaches to variance estimation.

\begin{table}[h!]
\caption{Comparison of standard errors computed under sampling based approaches (JK, BS, PB and WLBS) using a mixture of multivariate Gaussians model, using numerical derivatives and Hessians (\textbf{flexmix}), computed numerically via a Newton based method (NM), and computed analytically via the missing information principle (MIP). The data are from model M2, variable $2$.}
\label{tab:analyticSEs}
 \begin{center}
 \begin{tabular}{l|c|c|c|c|c}\hline
  &$\mathbf{\tau_1}$ & $\mathbf{\mu_1}$ & $\mathbf{\mu_2}$ & $\mathbf{\sigma^2_1}$ & $\mathbf{\sigma_2^2}$\\\hline
\bf{JK}   &0.015 &0.116 &0.029&0.037&0.012\\
\bf{BS}   &0.016 &0.126 &0.030&0.038&0.012\\
\bf{PB} &  0.017  & 0.109& 0.029 &0.027 & 0.012\\
\bf{WLBS} &0.016 &0.090 &0.031&0.018&0.013\\
\bf{flexmix} & 0.017 & 0.090 & 0.030 & 0.021 & 0.013 \\
\bf{NM}   &0.015 &0.096 &0.029&0.029&0.014\\
\bf{MIP}  &0.015 &0.096 &0.029&0.029&0.014\\ \hline
  \end{tabular}
 \end{center}
\end{table}

\subsubsection{Simulation Setting Two: $G = 3$.}
\label{sec:case two sim study results}

A second, similar simulation setting was used to analyze performance for a larger number of clusters i.e. $G = 3$ multivariate Gaussian components. Table $\ref{tab:sim case two tau results}$ presents the cluster probability estimates and associated standard errors, based on a single simulated data set under each of the models M5, M6, M7 and M8 with $n = 150$, for the JK, BS, PB and WLBS methods. The performance of the four approaches in the $G = 3$ multivariate Gaussian components case is similar to that summarized at the end of Section \ref{sec:case one sim study results}. Interestingly, the standard errors are larger for M5 (non-overlapping clusters) than for M6 (overlapping clusters), which upon examination was due to poor clustering performance for the particular data set simulated under M5.

\begin{table}[h]
\centering
\caption{Cluster probability estimates
(with associated standard errors) for a data set simulated under each of the four models detailed in Figure $\ref{fig:simulated data G3}$ where there are $G = 3$ multivariate Gaussian components, under each of the four variance estimation approaches. $\dagger$ For M5, $E_{BS} = 1017$  samples were drawn in order to successfully fit the model the requested number of times ($B_{BS} = 999$). $\ddagger$ $E_{JK} = 149$. $^* E_{BS} = 1838$. $^{**} E_{PB} = 1142$. $^{***} E_{BS} = 1002$. For M5 and M6 $\mathbf{\tau_{TRUE}}  = \{0.05, 0.05, 0.9\}$.  For M7 and M8 $\mathbf{\tau_{TRUE}}  = \{0.3, 0.3, 0.4\}$}
\vspace{0.5cm}
\label{tab:sim case two tau results}
\begin{tabular}{lll}\hline
& \multicolumn{1}{c}{$\mathbf{ \tau_{JK}} $} & \multicolumn{1}{c}{$\mathbf{ \tau_{BS} }$} \\
\textbf{M5}& \{0.15 (0.03), 0.15 (0.06), 0.70 (0.07)\} & \{0.15 (0.03), 0.15 (0.08), 0.70 (0.08)\}$^\dagger$ \\
\textbf{M6}& \{0.03 (0.01),  0.12 (0.03), 0.86 (0.03)\}$^\ddagger$ & \{0.03 (0.01), 0.12 (0.03), 0.85 (0.03)\}$^*$ \\
\textbf{M7}& \{0.29 (0.04), 0.32 (0.04), 0.39 (0.04)\} & \{0.29 (0.04), 0.32 (0.04), 0.39 (0.04)\} \\
\textbf{M8}& \{0.28 (0.13), 0.35 (0.07), 0.37 (0.09)\} & \{0.28 (0.08), 0.35 (0.07), 0.37 (0.06)\}$^{***}$ \\\hline
&  \multicolumn{1}{c}{$\mathbf{\tau_{PB}} $ }& \multicolumn{1}{c}{$ \mathbf{\tau_{WLBS} }$}\\
\textbf{M5}&   \{0.15 (0.03), 0.15 (0.08), 0.70 (0.08)\} & \{0.15 (0.03), 0.15 (0.08), 0.70 (0.08)\}\\
\textbf{M6}& \{0.03 (0.01), 0.12 (0.03), 0.85 (0.03)\}$^{**}$ &\{0.03 (0.01), 0.12 (0.03), 0.85 (0.03)\}\\
\textbf{M7}& \{0.29 (0.04), 0.32 (0.04), 0.39 (0.04)\} & \{0.29 (0.04), 0.32 (0.04), 0.39 (0.04)\}\\
\textbf{M8}&  \{0.28 (0.07), 0.35 (0.06), 0.37 (0.06)\} &\{0.28 (0.09), 0.35 (0.07), 0.37 (0.06)\}\\\hline
\end{tabular}
\end{table}

Again, $1000$ data sets were then generated under each of the models M5, M6, M7 and M8. For each data set an approximate $95$\% confidence interval was formed for each model parameter. Table $\ref{tab:tau1_sim_study_table 2}$ reports the number of data sets for which the confidence interval contained the true $\tau_1$ value. Clearly coverage performance is poorer in the $G = 3$ setting than in the $G = 2$ setting (Table $\ref{tab:tau1_sim_study_table 1}$), possibly attributable to the small number of observations ($n = 150$) and the challenging simulation parameter settings. The JK approach performs best in this context, particularly for models M5 and M6 where the other approaches struggle in terms of coverage performance. The WLBS performs slightly better than the BS and PB in terms of coverage. Both the BS and the PB commonly yield non-fitting issues, which do not affect the JK and WLBS approaches.

\begin{table}[!h]
\caption{Coverage results i.e. the proportion of data sets for which the true probability of membership of cluster one is within the JK, BS, PB and WLBS 95\% confidence intervals for the models M5, M6, M7 and M8, where there are $G = 3$ multivariate Gaussian components. The first column (`$\tilde{M}$ Fits') reports the number of simulated data sets to which it was possible to fit the optimal model $\tilde{M}$; a total of $1000$ data sets were simulated under each model setting. The `Non-fit' columns detail the average number of resamples that did not converge within each sampling procedure, with standard deviations given in parentheses. Note that in this study $B_{JK} = n = 150$ and $B_{BS} = B_{PB} = B_{WLBS} = 999$.}
\vspace{0.5cm}
\label{tab:tau1_sim_study_table 2}
\centering
\begin{tabular}{l|c|lc|lc|lc|lc|}
& $\tilde{M}$ & \multicolumn{2}{c|}{\bf{JK}} & \multicolumn{2}{c|}{\bf{BS}}&  \multicolumn{2}{c|}{\bf{PB}}&\multicolumn{2}{c|}{\bf{WLBS}}\\
& Fits & Non-fit & Coverage & Non-fit & Coverage & Non-fit & Coverage & Non-fit & Coverage\\\hline
\bf{M5} &  997 & 0.16 (2.15) & 0.616 & \,94.63 (398.59) & 0.117 & 15.38 (73.33) & 0.126 & 0.00 (0.00) & 0.166 \\
\bf{M6} & 996 & 0.05 (0.48) & 0.651 & 100.52 (355.51) & 0.276 & 23.55 (74.72) & 0.269 & 0.00 (0.00) & 0.331\\
\bf{M7} & 1000 & 0.00 (0.00) & 0.978 & 0.00 (0.00) & 0.974 & 0.00 (0.00) & 0.976 & 0.00 (0.00) & 0.967\\
\bf{M8} & 1000 & 0.00 (0.00) & 0.953 & \,1.05 (14.73) & 0.974 & 0.158 (3.22) & 0.959 & 0.00 (0.00) & 0.953\\
\end{tabular}
\end{table}

\subsection{Simulation Setting Three: higher dimensional scenarios.}
\label{sec:resultssim3}

To assess performance of the the JK, BS, PB and WLBS methods in higher dimensional scenarios, Table $\ref{tab:sim2 tau results}$ reports the cluster probability estimates and associated standard errors based on a single simulated data set under each of the settings for $n$, $p$ and $G = 5$ as described in Section \ref{sec:simdata3}. Note that for each setting $\tau_{TRUE} = (0.07, 0.07, 0.22, 0.27, 0.37)'$.

Adhering to the default settings in the \texttt{MclustBootstrap} function in \texttt{mclust}, $B_{BS} = B_{PB} = B_{WLBS} = 999$ samples are drawn for the BS and WLBS approaches respectively. By definition $B_{JK} = n$. In such higher dimensional settings the bootstrap in particular often has difficulty in successfully fitting a model to some of the $B_{BS} = 999$ samples drawn; this problem also occurs under the PB approach, but was never observed to occur under the WLBS approach as all observations are included (with some weight) in all samples. Thus Table $\ref{tab:sim2 tau results}$ also details the effective number of samples ($EB$) drawn for each approach.

\begin{table}[h]
\caption{Cluster probability estimates (with associated standard errors) for a data set from each of the simulation settings under each of the variance estimation approaches. The effective number of samples drawn ($EB$) is also reported. $\dagger$ The algorithm terminated as the number of non-estimable samples for model fitting reached the limit of $B_{BS} \times 10 = 9990$; only $40$ samples were successfully drawn and fitted and thus involved in the computation of the reported estimates. \vspace{0.5cm}}
\centering
\label{tab:sim2 tau results}
\begin{tabular}{|ll|l|}\hline
  & \bf{$n = 500$ \hspace{1cm} $p = 25$ \hspace{1cm} $G = 5$}   & $EB$ \\\hline
$ \mathbf{\tau_{JK}} $ &  \{0.06 (0.011), 0.08 (0.012), 0.24 (0.019), 0.26 (0.020), 0.36 (0.021)\} & 500\\
$ \mathbf{\tau_{BS}}$ & \{0.06 (0.009), 0.08 (0.009), 0.24 (0.020), 0.26 (0.018), 0.36 (0.018)\} & $\dagger$ \\
$ \mathbf{\tau_{PB}}$ & \{0.06 (0.009), 0.08 (0.012), 0.24 (0.018), 0.26 (0.020), 0.36 (0.021)\} & 1162 \\
$\mathbf{\tau_{WLBS}} $ & \{0.06 (0.010), 0.08 (0.012), 0.24 (0.019), 0.26 (0.020), 0.36 (0.022)\} & 999\\\hline
  & \bf{$n = 500$ \hspace{1cm} $p = 20$ \hspace{1cm} $G = 5$} & \\\hline
$ \mathbf{\tau_{JK}} $ &  \{0.06 (0.010), 0.08 (0.012), 0.24 (0.019), 0.26 (0.020), 0.36 (0.022)\} & 500 \\
$ \mathbf{\tau_{BS}} $ & \{0.06 (0.008), 0.08 (0.011), 0.24 (0.019), 0.26 (0.019), 0.36 (0.021)\} & 8142\\
$ \mathbf{\tau_{PB}} $ & \{0.06 (0.009), 0.08 (0.012), 0.24 (0.020), 0.26 (0.020), 0.36 (0.022)\} & 1069\\
$\mathbf{ \tau_{WLBS}} $ & \{0.06 (0.010), 0.08 (0.012), 0.24 (0.019), 0.26 (0.020), 0.36 (0.022)\} & 999\\\hline
& \bf{$n = 500$ \hspace{1cm} $p = 15$ \hspace{1cm} $G = 5$} & \\\hline
$ \mathbf{\tau_{JK}} $ &  \{0.07 (0.011), 0.07 (0.011), 0.23 (0.019), 0.25 (0.019), 0.38 (0.022)\} & 500 \\
$ \mathbf{\tau_{BS}} $ & \{0.07 (0.011), 0.07 (0.012), 0.23 (0.019), 0.25 (0.020), 0.38 (0.022)\} & 1025\\
$ \mathbf{\tau_{PB}} $ & \{0.07 (0.011), 0.07 (0.012), 0.23 (0.019), 0.25 (0.020), 0.38 (0.022)\} & 999\\
$\mathbf{ \tau_{WLBS}} $ & \{0.07 (0.011), 0.07 (0.012), 0.23 (0.019), 0.25 (0.020), 0.38 (0.022)\} & 999\\\hline\hline
& \bf{$n = 700$ \hspace{1cm} $p = 25$ \hspace{1cm} $G = 5$} & \\\hline
$ \mathbf{\tau_{JK}} $ &  \{0.06 (0.009), 0.08 (0.011), 0.22 (0.016), 0.28 (0.017), 0.36 (0.018)\} & 700 \\
$ \mathbf{\tau_{BS}} $ & \{0.06 (0.007), 0.08 (0.010), 0.22 (0.015), 0.28 (0.017), 0.36 (0.018)\} & 2125\\
$ \mathbf{\tau_{PB}} $ & \{0.06 (0.009), 0.08 (0.011), 0.22 (0.016), 0.28 (0.017), 0.36 (0.018)\} & 2076\\
$\mathbf{ \tau_{WLBS}} $ & \{0.06 (0.009), 0.08 (0.011), 0.22 (0.016), 0.28 (0.017), 0.36 (0.018)\} & 999\\\hline
& \bf{$n = 700$ \hspace{1cm} $p = 20$ \hspace{1cm} $G = 5$} & \\\hline
$ \mathbf{\tau_{JK}} $ &  \{0.06 (0.009), 0.09 (0.011), 0.22 (0.016), 0.27 (0.017), 0.36 (0.018)\} & 700 \\
$ \mathbf{\tau_{BS}} $ & \{0.06 (0.008), 0.09 (0.011), 0.22 (0.016), 0.27 (0.016), 0.36 (0.018)\} & 1052\\
$ \mathbf{\tau_{PB}} $ & \{0.06 (0.009), 0.09 (0.010), 0.22 (0.016), 0.27 (0.016), 0.36 (0.018)\} & 1022\\
$\mathbf{ \tau_{WLBS}} $ & \{0.06 (0.009), 0.09 (0.011), 0.22 (0.015), 0.27 (0.016), 0.36 (0.018)\} & 999\\\hline
& \bf{$n = 700$ \hspace{1cm} $p = 15$ \hspace{1cm} $G = 5$} & \\\hline
$ \mathbf{\tau_{JK}} $ &  \{0.07 (0.009), 0.08 (0.01), 0.23 (0.016), 0.25 (0.016), 0.37 (0.018)\} & 700 \\
$ \mathbf{\tau_{BS}} $ & \{0.07 (0.010), 0.08 (0.01), 0.23 (0.016), 0.25 (0.017), 0.37 (0.019)\} & 1059 \\
$ \mathbf{\tau_{PB}} $ & \{0.07 (0.009), 0.08 (0.011), 0.23 (0.016), 0.25 (0.017), 0.37 (0.018)\} & 1025 \\
$\mathbf{ \tau_{WLBS}} $ & \{0.07 (0.009), 0.08 (0.01), 0.23 (0.015), 0.25 (0.015), 0.37 (0.018)\} & 999\\\hline
\end{tabular}
\end{table}

To assess the computational performance of the methods in more complex scenarios, $100$ data sets were generated under each of the simulation settings for $n, p$ and $G$. For each data set the run time for each method was recorded and summaries are reported in Table $\ref{tab:runtimes_sim_study2}$. Across all high dimensional settings the JK is the cheapest computationally, however both it and the BS are prone to model fitting issues, meaning the final variance estimates produced are not always based on the number of samples expected or requested by the user. This phenomenon occurs more frequently in settings where the $n/p$ ratio is small. The WLBS and PB are markedly more stable in terms of model fitting, though they are slower than the JK and BS methods. That the WLBS does not often encounter fitting issues is due to the fact that the same data set used to estimate $\tilde{M}$ is used in the WLBS procedure. The large variance of $80.67$ for the $n = 500$, $p = 20$ WLBS setting in Table $\ref{tab:runtimes_sim_study2}$ is due to one very large run time. When this runtime was omitted the mean and standard deviation are $44.05$ $(2.73)$; the run times for the BS and WLBS for this isolated simulated data set were also relatively large.

Also included for comparative purposes in Table $\ref{tab:runtimes_sim_study2}$ are summaries of the run times taken to compute the standard errors from a version of the information matrix following the empirical Fisher information standard error formula as recommended in \cite{boldea2009}; in all cases these run times are notably larger than those from the sampling based approaches. In terms of estimates, for example from an $n = 500$ and $p = 25$ simulated data set, the Boldea \& Magnus approach estimates the mixing probabilities and associated standard errors to be $\tau = (0.07 (0.005), 0.06 (0.009), 0.24 (0.009), 0.27 (0.010), 0.35 (0.011))$, which are not notably different to those reported in Table \ref{tab:sim2 tau results}.

\begin{table}[!h]
\caption{Average run times in seconds (standard deviations in parentheses) for different simulation settings in high dimensional scenarios. In all settings $G = 5$. The third column ($\tilde{M}$ Fits) details the number of the 100 simulated data sets for which it was possible to fit the optimal model $\tilde{M}$. Under each of the JK, BS, PB, WLBS and Boldea \& Magnus headings the second column (Fits) details the number of the $\tilde{M}$ Fits data sets for which the effective number of samples $EB$ was equal to that requested i.e. equal to $B_{JK} = n$ and $B_{BS} = B_{PB} = B_{WLBS} = 999$. }
\label{tab:runtimes_sim_study2}
\centering
\begin{tabular}{|cc|c|lc|lc|lc|lc|lc|}\hline
 &  & $\mathbf{\tilde{M}}$ &  \multicolumn{2}{|c|}{\bf{JK}}& \multicolumn{2}{|c|}{\bf{BS}} & \multicolumn{2}{|c|}{\bf{PB}} & \multicolumn{2}{|c|}{\bf{WLBS}} & \multicolumn{2}{|c|}{\bf{Bol/Mag}} \\
$\mathbf{n}$ & $\mathbf{p}$ & \bf{Fits} &  Time & Fits & Time & Fits & Time & Fits & Time & Fits & Time & Fits\\\hline
500 & 25 & 90 & 5.2 (0.09) & 84 &  38.5 (5.65) & 13 & 37.2 (8.12) & 90 & 61.6 (2.9) & 90 & 648.9 (3.98) &  90 \\
500 & 20 & 99 & 4.2 (0.26) & 96 & 21.7 (10.63) & 69 & 21.9 (3.67) & 99 & 52.2 (80.67) & 99 & 192.6 (7.26) & 99 \\
500 & 15 & 100 & 3.1 (0.11) & 99 & 7.1 (2.74) & 99 & 15.5 (1.08) & 100 & 30.3 (0.81) & 100 & 42.1 (0.6) & 100\\\hline
700 & 25 & 100 & 10.5 (0.12) & 100  & 22.7 (12.94) & 94 & 64.9 (30.97)  & 100 & 72.3 (0.41) & 100 & 901.3 (7.39) & 100\\
700 & 20 & 100 & 8.0 (0.17) & 100 & 11.6 (0.80)  &  100 & 28.4 (5.80) & 100 & 51.3 (1.06) & 100 & 270.3 (0.70) & 100 \\
700 & 15 & 100 & 5.8 (0.17) & 100 & 8.2 (0.16) & 100 & 16.5 (1.91)  & 100 & 35.4 (0.53) & 100 & 60.8 (0.35) & 100 \\\hline
\end{tabular}
\end{table}

\subsection{\emph{Old Faithful} results}
\label{sec:faithful results}

While Section $\ref{sec:sim study results}$ demonstrated the advantages and disadvantages of the sampling based methods through a simulation study, here the utility of the methods is illustrated through a real clustering problem where true parameter estimates are unknown. For the \emph{Old Faithful} data, under \texttt{mclust}, the optimal mixture of Gaussians model has $G = 3$ components and common covariance structure $\bSigma_g =\bSigma$ across groups, based on BIC. JK, BS, PB and WLBS parameter estimates and associated standard errors for this model are presented below, along with cluster covariance estimated values (with associated standard errors). The maximum likelihood parameter estimates found using the single best \texttt{mclust} model based on the full data are also included for comparative purposes in both cases.

\begin{center}
\begin{tabular}{ll}
$ \tau_{MCLUST}$ & = $\left(\begin{array}{ccc} 0.46, & 0.36, & 0.18  \end{array}\right)$ \\
$ \tau_{JK}$ & = $\left(\begin{array}{ccc} 0.46 \: (0.04), & 0.36 \: (0.03), & 0.18 \: (0.04)  \end{array}\right)$ \\
$ \tau_{BS}$ & = $\left(\begin{array}{ccc} 0.47 \: (0.05), & 0.36 \: (0.03), & 0.17 \: (0.05)  \end{array}\right)$ \\
$ \tau_{PB}$ & = $\left(\begin{array}{ccc} 0.48 \: (0.04), & 0.36 \: (0.03), & 0.16 \: (0.03)  \end{array}\right)$ \\
$ \tau_{WLBS}$ & = $\left(\begin{array}{ccc} 0.48 \: (0.06), & 0.36 \: (0.03), & 0.16 \: (0.05)  \end{array}\right)$ \\
 & \\
 & \\
$ \mu_{MCLUST}$ & = $\left(\begin{array}{ccc}4.48, & 2.04, & 3.82 \\ 80.89, & 54.49, & 77.65\\\end{array}\right)$\\&\\
$ \mu_{JK}$ & = $\left(\begin{array}{ccc}4.47 \: (0.03), & 2.04 \: (0.03), & 3.81 \: (0.06), \\ 80.89 \: (0.47), & 54.49 \: (0.60), & 77.62 \: (1.18)\\\end{array}\right)$\\
                                     & \\
$ \mu_{BS}$ & = $\left(\begin{array}{ccc} 4.47 \: (0.05), & 2.03 \: (0.03), & 3.79 \: (0.11)\\ 80.86 \: (0.59), & 54.45 \: (0.59), & 77.37 \: (2.24)\\\end{array}\right)$\\
  & \\
$ \mu_{PB}$ & = $\left(\begin{array}{ccc} 4.47 \: (0.03), & 2.04 \: (0.03), & 3.79 \: (0.06)\\ 80.84 \: (0.60), & 54.49 \: (0.59), & 77.49 \: (1.16)\\\end{array}\right)$\\
  & \\
$ \mu_{WLBS}$ & = $\left(\begin{array}{ccc} 4.46 \: (0.05), & 2.03 \: (0.03), & 3.76 \: (0.13)\\ 80.81 \: (0.59), & 54.44 \: (0.61), & 76.97 \: (2.41)\\\end{array}\right)$\\
\end{tabular}
\end{center}

\vspace{5mm}
\begin{center}
\begin{tabular}{llllll}
\hspace{-2mm}$\Sigma_{MCLUST}$ &
$\hspace{-4mm}= \hspace{-1mm}\left(\begin{array}{cc}
0.08 &  0.48 \\
0.47 &  33.74 \end{array}\right)$
 &
$\Sigma_{JK}$ &
$\hspace{-4mm}= \hspace{-1mm}\left(\begin{array}{cc}
0.08 \:(0.01) &       0.47   \:(0.12)\\
0.47  \:(0.12)&      33.73  \:(2.77) \end{array}\right)$
 &
$\Sigma_{BS}$ &
$\hspace{-4mm}= \hspace{-1mm}\left(\begin{array}{cc}
0.08 \:(0.01) &       0.46   \:(0.15)\\
0.46  \:(0.15)&      32.88  \:(2.83) \end{array}\right)$
\end{tabular}

\vspace{2mm}

\begin{tabular}{llll}
$\Sigma_{PB}$ &
$\hspace{-4mm}= \hspace{-1mm}\left(\begin{array}{cc}
0.08 \:(0.01) &       0.48   \:(0.12)\\
0.48  \:(0.12)&      33.32  \:(2.76) \end{array}\right)$
&
$\Sigma_{WLBS}$ &
$\hspace{-4mm}= \hspace{-1mm}\left(\begin{array}{cc}
0.08 \:(0.01) &       0.45   \:(0.16)\\
0.45  \:(0.16)&      32.94  \:(2.89) \end{array}\right)$
\end{tabular}
\end{center}

The standard errors for all parameters under each method are small relative to the size of the parameter estimates themselves. The standard errors using the BS and WLBS are slightly larger than their JK and PB counterparts for most parameters. This is to be expected as there is likely to be much less variability in the estimates arising from the JK samples than would be observed in the BS or WLBS cases, as each JK sample differs only by one observation. On the other hand the BS and WLBS samples are likely to differ from each other to a greater degree. Similar results have been presented previously for this data set in a univariate context \citep[][page 139--155]{everitt09}.

\begin{figure}
\begin{center}
\begin{tabular}{cc}
\subfigure[]{\label{fig:Fig5a}\includegraphics[width=6cm,height=7cm]{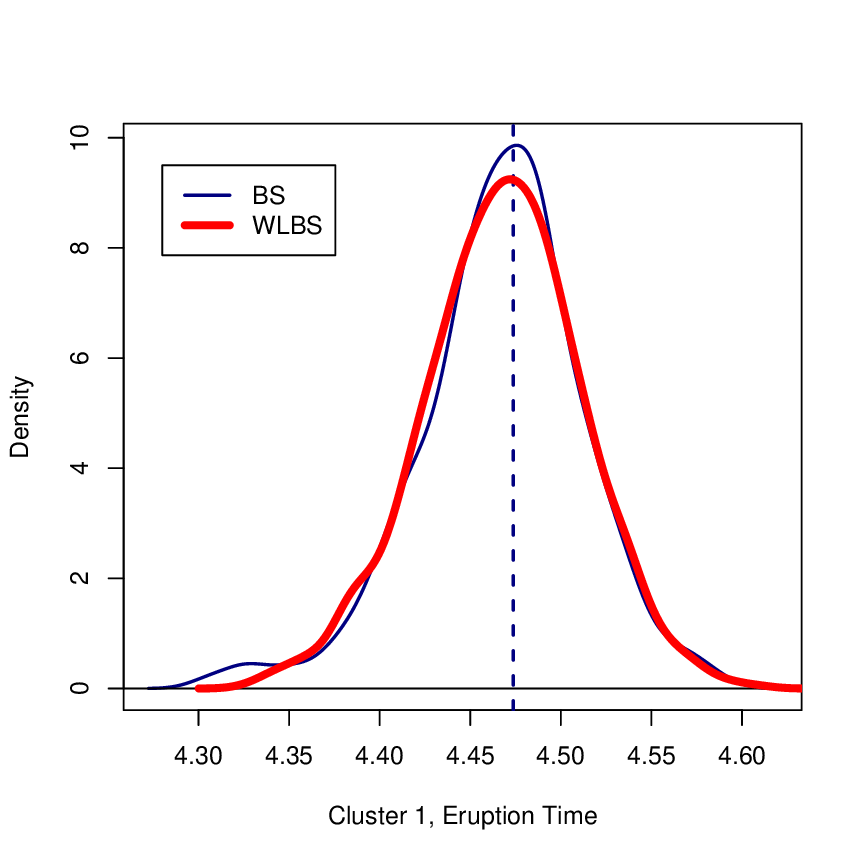}}&
\subfigure[]{\label{fig:Fig5b}\includegraphics[width=6cm,height=7cm]{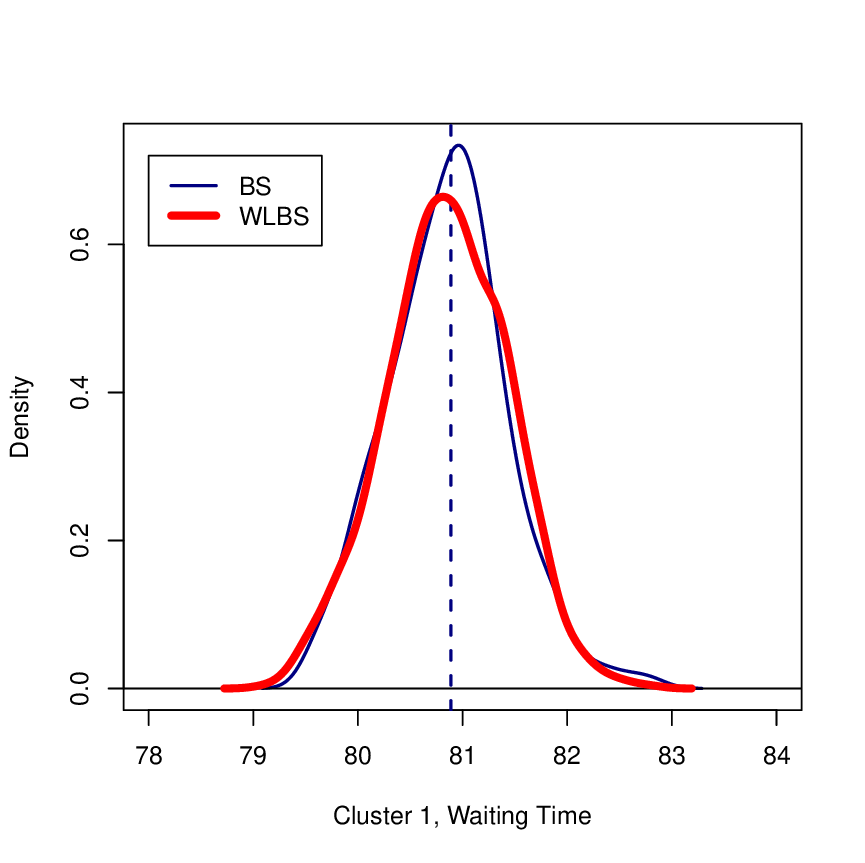}}\\
\subfigure[]{\label{fig:Fig5c}\includegraphics[width=6cm,height=7cm]{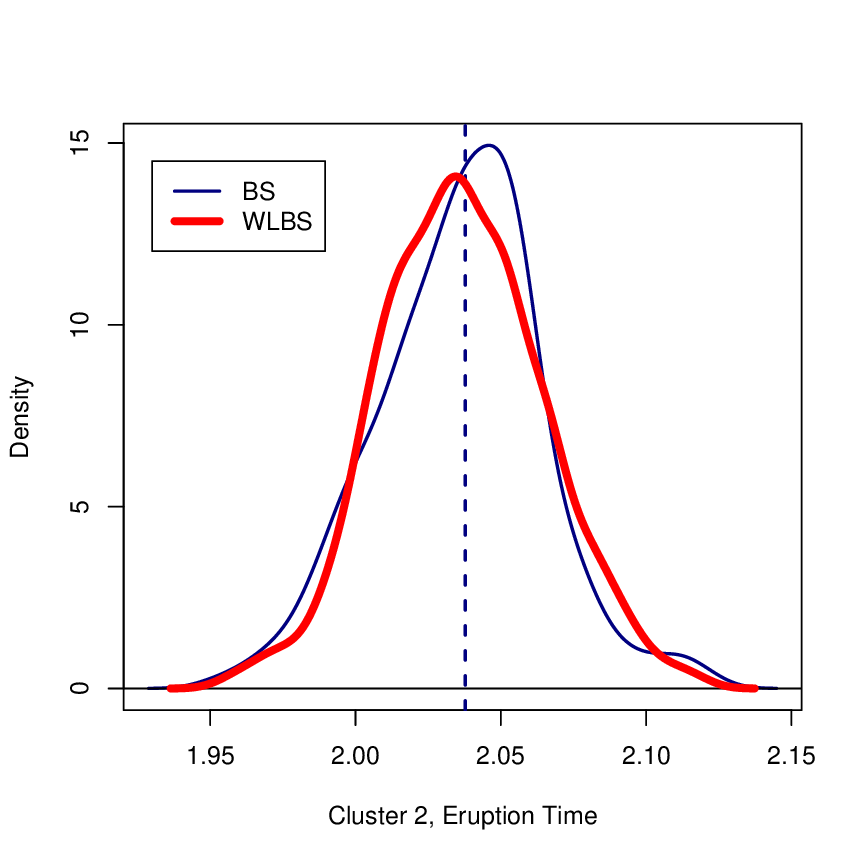}}&
\subfigure[]{\label{fig:Fig5d}\includegraphics[width=6cm,height=7cm]{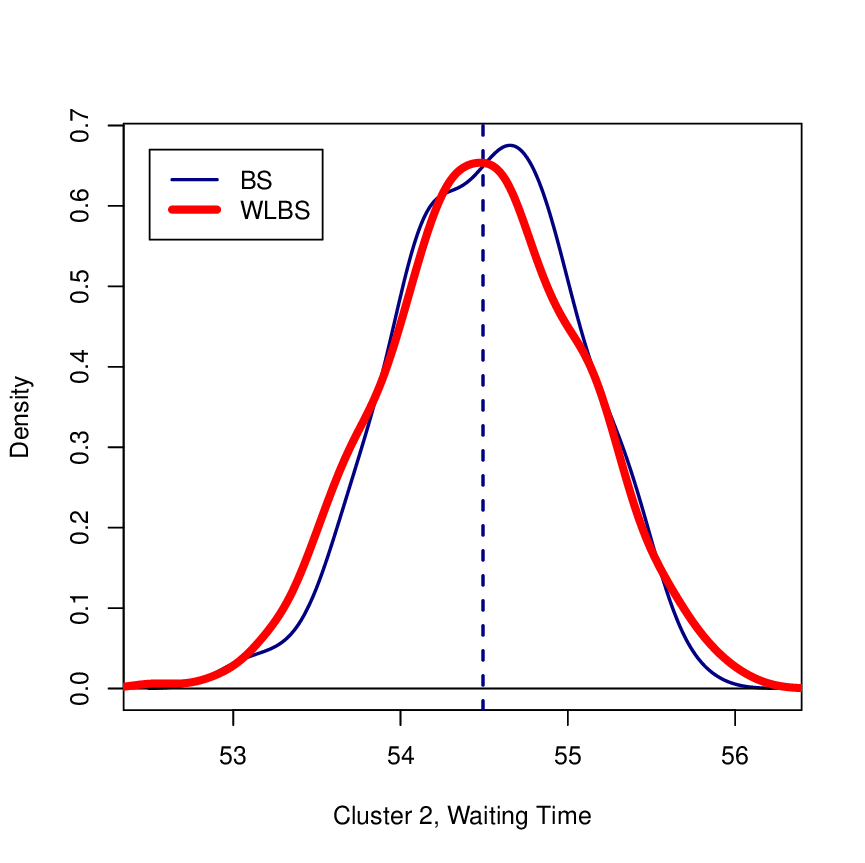}}\\
\subfigure[]{\label{fig:Fig5e}\includegraphics[width=6cm,height=7cm]{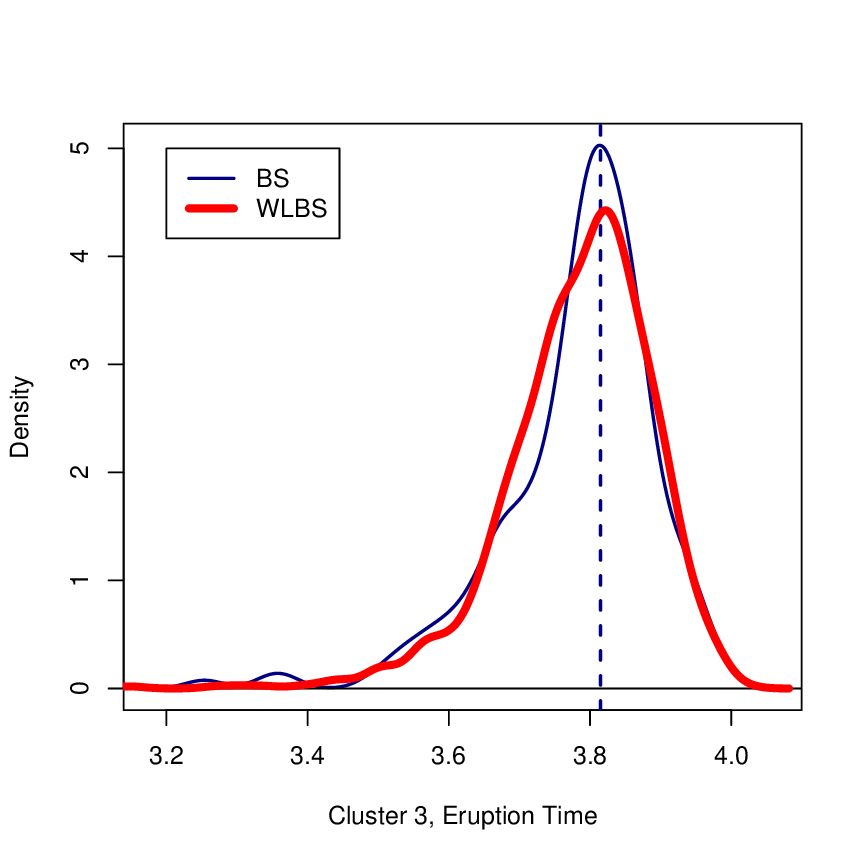}}&
\subfigure[]{\label{fig:Fig5f}\includegraphics[width=6cm,height=7cm]{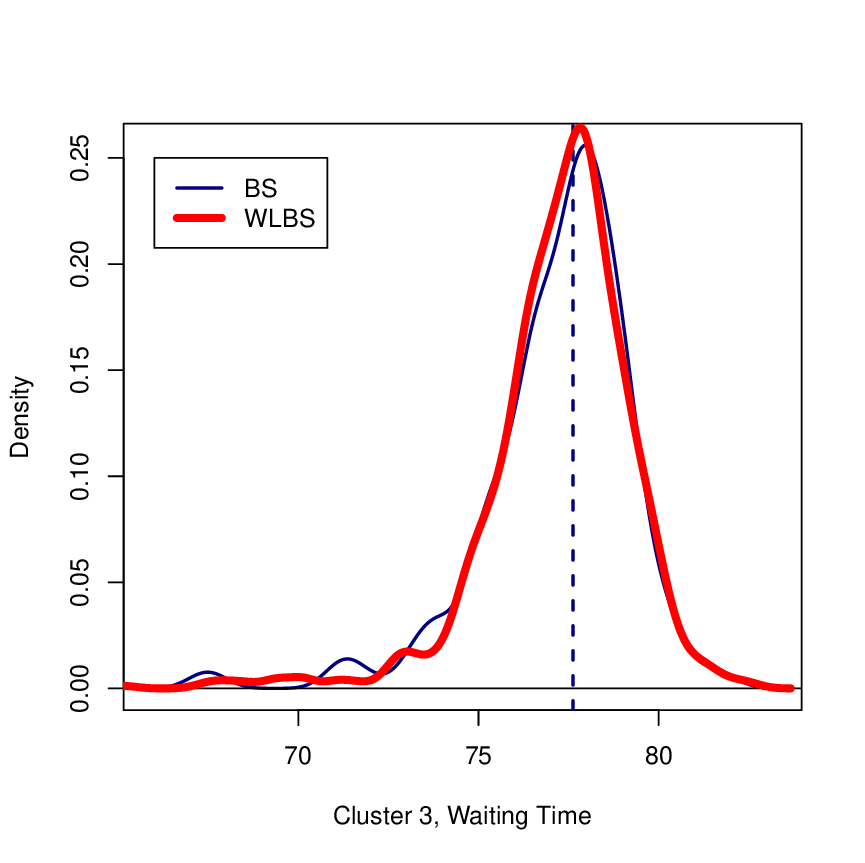}}\\
\end{tabular}
\caption{Kernel density plots of the BS and WLBS estimates of the
$\bmu$ parameters for the \emph{Old Faithful} data. The solid blue line and the thicker solid red line represent the BS and WLBS kernel
densities respectively. The dashed lines represent the values of the MLEs from the model fitted to the full data set. The PB kernel densities are extremely close to the BS lines and for visual clarity have been omitted.}
\label{fig:Faithful_mu_BS WLBS_plots_2}
\end{center}
\end{figure}

\begin{figure}
\begin{center}
\begin{tabular}{c}
\subfigure[]{\label{fig:Fig6a}\includegraphics[width=8cm,height=7cm]{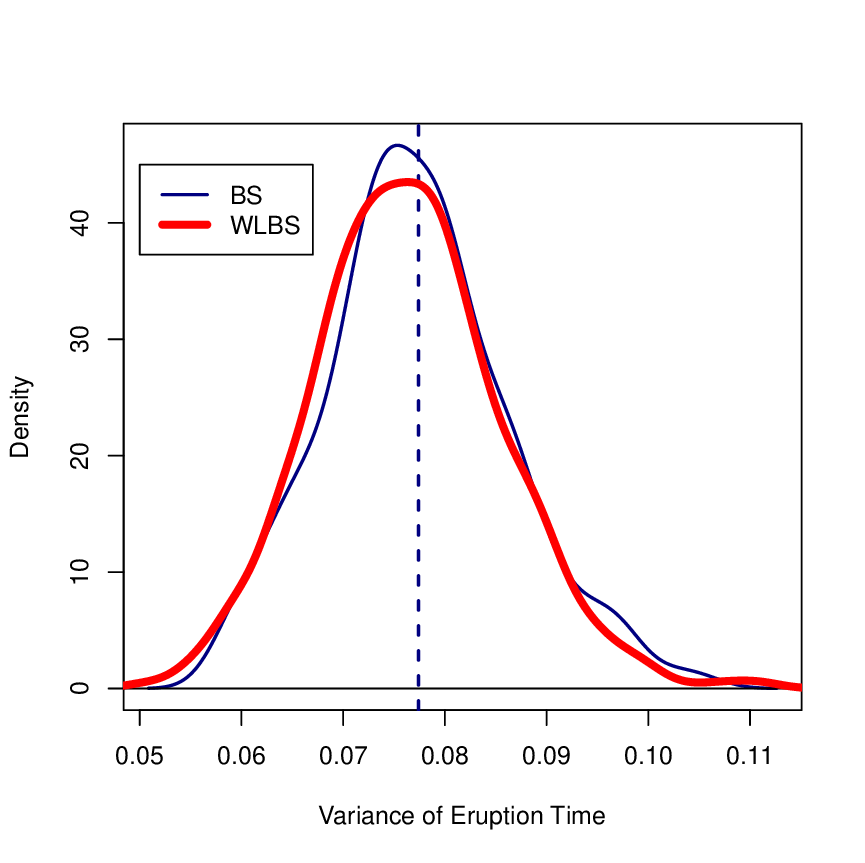}}\\
\subfigure[]{\label{fig:Fig6b}\includegraphics[width=8cm,height=7cm]{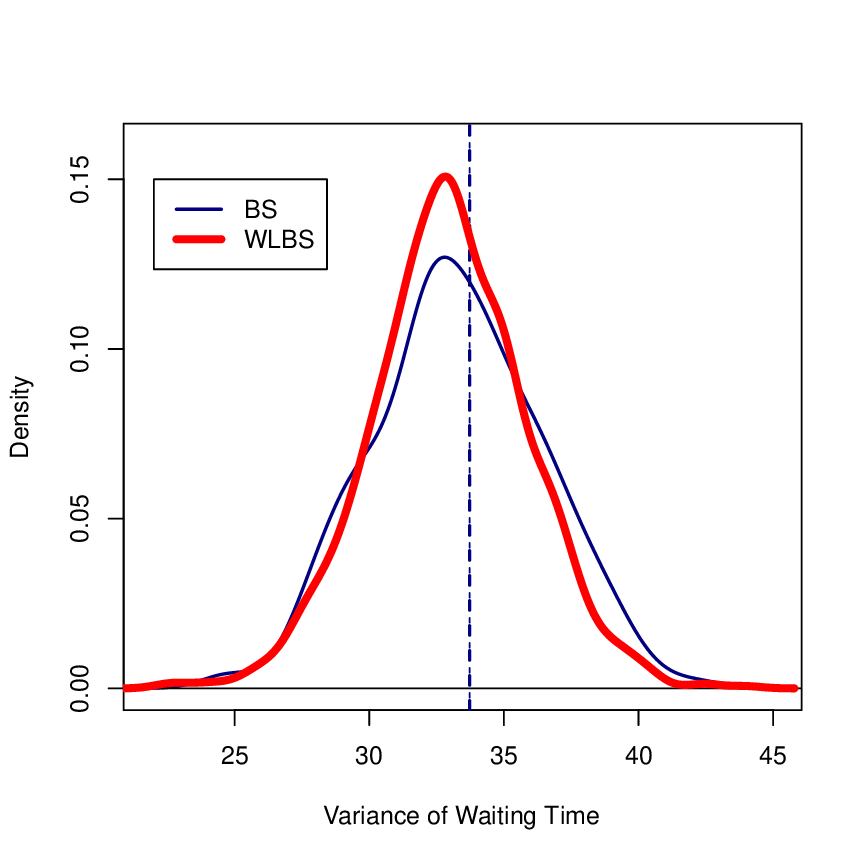}}\\
\subfigure[]{\label{fig:Fig6c}\includegraphics[width=8cm,height=7cm]{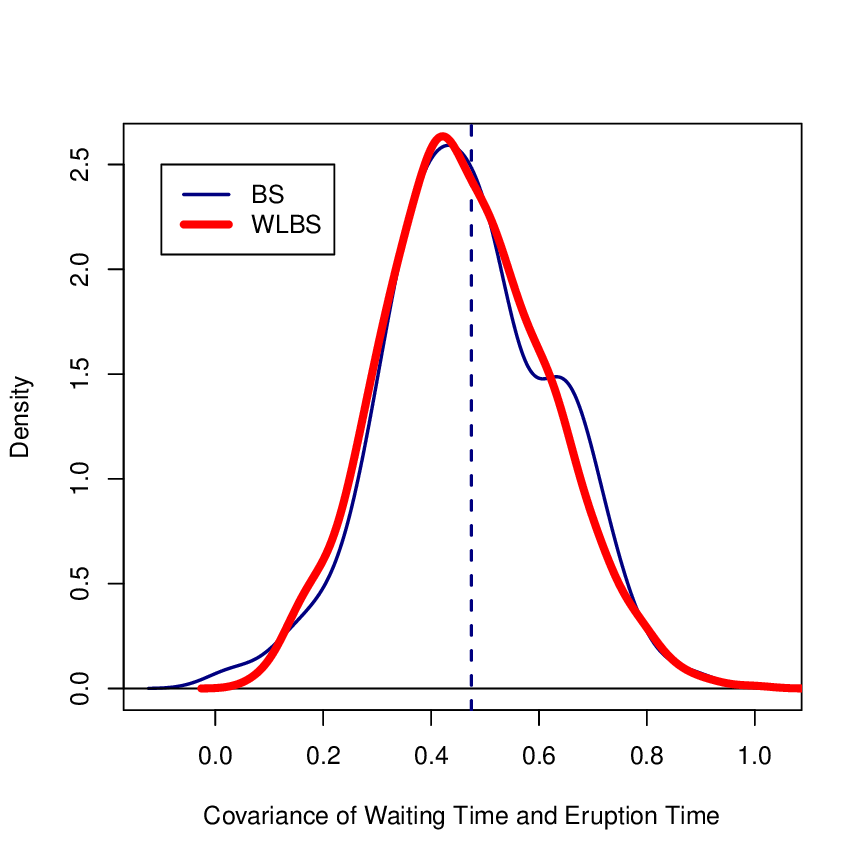}}\\
\end{tabular}
\caption{Kernel density plots of the BS and WLBS estimates of the
$\bSigma$ parameters for the \emph{Old Faithful} data. The solid blue line and the thicker solid red line represent the BS and WLBS kernel
densities respectively. The dashed lines represent the values of the MLEs from the model fitted to the full data set. The PB kernel densities are extremely close to the BS lines and for visual clarity have been omitted.}
\label{fig:Faithful_sigma_BS_WLBS_plots_2}
\end{center}
\end{figure}

The sampling based approaches to variance estimation discussed provide not only estimates of the model parameters, but also insight as to their associated uncertainty, which can be graphically illustrated. Figure $\ref{fig:Faithful_mu_BS WLBS_plots_2}$ provides kernel density plots for the mean waiting duration and eruption duration for all three clusters. The plots indicate good agreement between the BS and WLBS approaches; notably the WLBS densities are flatter in some cases. In addition, the kernel density plots for the model's covariance parameters are provided in Figure $\ref{fig:Faithful_sigma_BS_WLBS_plots_2}$. 
Plotting the JK density estimates for the model parameters results in very bumpy and very narrow densities. This is due to the similarity of the JK samples and therefore the parameter estimates themselves (necessitating the use of pseudo-values in computing confidence intervals for the JK approach). The PB kernel density results were also computed but are extremely close to the outcomes for the BS in the case of Figures $\ref{fig:Faithful_mu_BS WLBS_plots_2}$ and $\ref{fig:Faithful_sigma_BS_WLBS_plots_2}$ and hence for visual clarity have been omitted from the plots.


%

\subsection{\emph{Thyroid} results}
\label{sec:thyroid results}

For the \emph{Thyroid} data, under \texttt{mclust}, the optimal mixture of Gaussians model has $G = 3$ components and diagonal covariance structure with varying volume and shape, $\bSigma_g =\lambda_g A_g$, across groups.
The results for the estimation of parameters and their associated uncertainties under the optimal model are presented below using $3$ multivariate Gaussian components and covariance structure $\bSigma_g =\lambda_g A_g$. The maximum likelihood parameter estimates found using the single best \texttt{mclust} model based on BIC are also included for comparative purposes. Covariance parameter estimates and associated standard errors are detailed in Appendix \ref{app:thyroid}, along with the \textbf{MclustBootstrap} code used to obtain the results.

\begin{center}
\begin{tabular}{ll}
$ \tau_{MCLUST}$ & = $\left(\begin{array}{ccc} 0.71, & 0.16, & 0.13  \end{array}\right)$ \\
$ \tau_{JK}$ & = $\left(\begin{array}{ccc} 0.74 \: (0.03), & 0.15 \: (0.03), & 0.11 \: (0.02)  \end{array}\right)$ \\
$ \tau_{BS}$ & = $\left(\begin{array}{ccc} 0.71 \: (0.03), & 0.16 \: (0.03), & 0.13 \: (0.02)  \end{array}\right)$ \\
$ \tau_{PB}$ & = $\left(\begin{array}{ccc} 0.71 \: (0.03), & 0.16 \: (0.03), & 0.13 \: (0.02)  \end{array}\right)$ \\
$ \tau_{WLBS}$ & = $\left(\begin{array}{ccc} 0.71 \: (0.03), & 0.16 \: (0.03), & 0.13 \: (0.02)  \end{array}\right)$ \\
\end{tabular}
\end{center}

\begin{center}
\begin{tabular}{ll}
$ \mu_{MCLUST}$ & = $\left(\begin{array}{ccccc} 110.34, & 9.09, & 1.72, & 1.31, & 2.49 \\ 95.53, & 17.69, & 4.27, & 0.97, & -0.02 \\ 123.22, & 3.79, & 1.06, & 13.91, & 18.84 \\ \end{array}\right)$\\
\\
$ \mu_{JK}$ & = $\left(\begin{array}{ccccc} 110.13 \: (0.66),& 9.11 \: (0.18), & 1.74 \: (0.04), & 1.35 \: (0.04), & 2.42 \: (0.15)\\ 95.05 \: (3.27), & 18.40 \: (0.77), & 4.46 \: (0.40), & 0.96 \: (0.07), & 0.00 \: (0.05)\\ 124.46 \: (1.96), & 3.32 \: (0.45), & 0.96 \: (0.11), & 15.03 \: (2.44), & 20.71 \: (2.99)\\ \end{array}\right)$\\
& \\
$ \mu_{BS}$ & = $\left(\begin{array}{ccccc} 110.33 \: (0.65),& 9.09 \: (0.19), & 1.72 \: (0.04), & 1.31 \: (0.04), & 2.50 \: (0.15)\\ 95.59 \: (3.37), & 17.68 \: (0.81), & 4.28\: (0.43), & 0.97 \: (0.07), & -0.02 \: (0.05)\\ 123.36 \: (0.68), & 3.74 \: (0.19), & 1.05 \: (0.04), & 14.12 \: (0.04), & 18.97 \: (0.15)\\ \end{array}\right)$\\
& \\
$ \mu_{PB}$ & = $\left(\begin{array}{ccccc}
110.35 \: (0.65),& 9.08 \: (0.18), & 1.72 \: (0.04), & 1.31 \: (0.04), & 2.50 \: (0.15)\\
95.44 \: (3.20), & 17.68 \: (0.75), & 4.25\: (0.40), & 0.97 \: (0.07), & -0.02 \: (0.05)\\
123.27 \: (1.92), & 3.79 \: (0.40), & 1.06 \: (0.10), & 13.88 \: (2.35), & 18.87 \: (2.88)\\ \end{array}\right)$\\
& \\
$ \mu_{WLBS}$ & = $\left(\begin{array}{ccccc} 110.34 \: (0.68), & 9.10 \: (0.19), & 1.72 \: (0.04), & 1.31 \: (0.04), & 2.50 \: (0.15) \\ 95.42 \: (3.38), & 17.68 \: (0.82), & 4.28 \: (0.42), & 0.97 \: (0.07), & -0.01 \: (0.05)\\ 123.37 \: (1.83), & 3.73 \: (0.40), & 1.04 \: (0.10), & 13.96 \: (2.37), & 18.87 \: (2.75)\\ \end{array}\right)$\\
\end{tabular}
\end{center}

The standard errors for all parameters under each method are small relative to the size of the parameter estimates themselves and are approximately equal across the variance estimation approaches considered. The parameter estimates across all methods are close to those from the full data model, with the PB and WLBS proving most accurate in this regard. This verifies that the JK, BS, PB and WLBS approaches presented are robust even in this higher dimensional real data application and that there is evidence to favour adoption of the PB or WLBS approaches if a single method is to be preferred.

From a computational perspective, all variance estimation approaches are computationally efficient. The results produced throughout Section \ref{sec:results} were obtained via the \texttt{MclustBootstrap} function in the most recent version of the \textbf{R} package \texttt{mclust}. In the context of clustering the \emph{Old Faithful} data, the JK, BS, PB and WLBS approaches to variance estimation required $0.19$s, $2.45$s, $5.33$s and $70.47$s to run respectively on a $2.8$ GHz Mac OS X laptop, where the default $999$ samples were requested in the BS, PB and WLBS settings. The corresponding times for the \emph{Thyroid} data set were $0.13$s, $0.77$s, $4.03$s and $15.32$s. The increase in the WLBS setting over the JK, BS and PB settings for computational time is due to the required maximization of the weighted complete data likelihood, but the overall computational cost is still cheap from a user perspective. The JK and BS methods are in turn faster than the PB due to the time required to simulate from the model under the latter approach. The \textbf{MclustBootstrap} code used to obtain the \emph{Thyroid} results is provided at the end of Appendix \ref{app:thyroid}.

\section{Discussion and further work}
\label{sec:discussion}

Although model-based clustering is now a widely used approach to clustering in a range of disciplines, especially through the use of the \texttt{mclust} package in \textbf{R}, little attention has previously been paid to providing estimates of the variance associated with parameter estimates. Here, four sampling based approaches to variance estimation are discussed in the context of model-based clustering. The jackknife, bootstrap and parametric bootstrap approaches to variance estimation are basic tools in any statistician's toolkit, but difficulties with the bootstrap in particular arise in the clustering context when small clusters are present. The weighted likelihood bootstrap addresses this shortcoming. The WLBS has been shown to perform as well as the JK, BS and PB in general, and particularly well in the presence of small clusters. In terms of which sampling based approach the practitioner should use to obtain variance estimates, the simulation studies and real applications presented here suggest that when roughly similarly sized clusters are present, all four methods perform equally well. In the presence of small clusters or in high dimensional settings however, the JK and WLBS are more stable than the BS and PB. Although it comes with higher, yet still user friendly, computational cost, overall the WLBS is found to be the preferred method of variance estimation. This is primarily because, if the model provides a poor fit, the BS and PB will either fail and/or require extra samples, and the jackknife samples may not well represent the full data set, whereas the WLBS will give weight to all observations and provide a solution. This tends to occur in cases of small and/or overlapping clusters. However, the poor model fit is attributable to the nature of the data and not the fault of the sampling variance estimation approaches themselves. Indeed, instances of the BS and PB needing extra samples, or failing, may in fact be evidence of poor model fit in the first instance, providing an additional diagnostic tool in this regard.

The standard errors calculated from the JK, BS, PB and WLBS have several practical uses including formation of approximate confidence intervals for parameter estimates, construction of hypothesis tests as to whether parameters should be included in the model and analysis of the bias of maximum
likelihood parameter estimates versus their JK/BS/PB/WLBS counterparts as a means of assessing model goodness of fit. In a similar vein, the sampling based methods discussed could be employed as an aid to model selection.

Further avenues of research are plentiful and varied. For example, an application that would perhaps be of interest to an \texttt{mclust} user would be the quantification of the standard errors of the parameters constituting the eigenvalue decomposition of the covariance matrix into its size, orientation and shape components, $\Sigma_g = \lambda_g D_g A_g D_g^T$. This may aid in the process of model specification, namely in determining which parameters could be set equal across groups to achieve a more parsimonious decomposition. Specific to the weighted likelihood bootstrap method, a  more thorough investigation of alternative Dirichlet parameterisations or alternative weighting distributions could be conducted to examine their stability and suitability in settings where some clusters are sparsely populated. In addition, the JK, BS, PB and WLBS could be examined in the context of non-Gaussian mixtures, such as mixtures of $t$ distributions or skew-$t$ distributions \citep{lee2013emmix,lee2013model}.

\newpage
\bibliographystyle{spbasic}      
\bibliography{AOH_JK_BS_paper}   

\begin{thebibliography}{48}
\providecommand{\natexlab}[1]{#1}
\providecommand{\url}[1]{{#1}}
\providecommand{\urlprefix}{URL }
\expandafter\ifx\csname urlstyle\endcsname\relax
  \providecommand{\doi}[1]{DOI~\discretionary{}{}{}#1}\else
  \providecommand{\doi}{DOI~\discretionary{}{}{}\begingroup
  \urlstyle{rm}\Url}\fi
\providecommand{\eprint}[2][]{\url{#2}}

\bibitem[{Andrews and Buchinsky(2000)}]{andrews2000three}
Andrews DW, Buchinsky M (2000) A three-step method for choosing the number of
  bootstrap repetitions. Econometrica 68(1):23--51

\bibitem[{Andrews and Guggenberger(2009)}]{AndrewGuggenberger2008}
Andrews DW, Guggenberger P (2009) Incorrect asymptotic size of subsampling
  procedures based on post-consistent model selection estimators. Journal of
  Econometrics 152(1):19--27

\bibitem[{Azzalini and Bowman(1990)}]{azzalini1990}
Azzalini A, Bowman A (1990) A look at some data on the {O}ld {F}aithful geyser.
  Applied Statistics 39(3):357--365

\bibitem[{Basford et~al(1997)Basford, Greenway, McLachlan, and
  Peel}]{basford97}
Basford K, Greenway D, McLachlan G, Peel D (1997) Standard errors of fitted
  means under normal mixture models. Computational Statistics 12:1--17

\bibitem[{Boldea and Magnus(2009)}]{boldea2009}
Boldea O, Magnus J (2009) Maximum likelihood estimation of the multivariate
  normal mixture model. Journal of the American Statistical Association
  104:1539--1549

\bibitem[{B{\"u}hlmann(1997)}]{buhlmann1997sieve}
B{\"u}hlmann P (1997) Sieve bootstrap for time series. Bernoulli 3(2):123--148

\bibitem[{Coomans et~al(1983)Coomans, Broeckaert, Jonckheer, and
  Massart}]{coomans1983comparison}
Coomans D, Broeckaert I, Jonckheer M, Massart D (1983) Comparison of
  multivariate discrimination techniques for clinical data—application to the
  thyroid functional state. Methods of Information in Medicine 22(02):93--101

\bibitem[{Davison and Hinkley(1997)}]{davison1997bootstrap}
Davison AC, Hinkley DV (1997) Bootstrap methods and their application, vol~1.
  Cambridge university press, Cambridge

\bibitem[{Dempster et~al(1977)Dempster, Laird, and Rubin}]{dempster77}
Dempster A, Laird N, Rubin D (1977) Maximum likelihood from incomplete data via
  the {EM} algorithm. Journal of the Royal Statistical Society Series B
  (Methodological) 39(1):1--38

\bibitem[{Diebolt and Ip(1996)}]{diebolt96}
Diebolt J, Ip E (1996) Stochastic {EM}: method and application. In: WR~Gilks
  SR, Spiegelhalter D (eds) Markov Chain Monte Carlo in Practice, Chapman \&
  Hall, London, pp 259--273

\bibitem[{Efron(1981)}]{efron81_2}
Efron B (1981) Nonparametric estimates of standard error: The jackknife, the
  bootstrap and other methods. Biometrika 68(3):589--599

\bibitem[{Efron(1982)}]{efron1982}
Efron B (1982) The jackknife, the bootstrap, and other resampling plans,
  vol~38. Siam, Philadelphia

\bibitem[{Efron(1994)}]{efron94}
Efron B (1994) Missing data, imputation and the bootstrap (with discussion).
  Journal of the American Statistical Association 89(426):463--479

\bibitem[{Efron and Stein(1981)}]{efron81}
Efron B, Stein C (1981) The jackknife estimate of variance. The Annals of
  Statistics 9(3):586--596

\bibitem[{Efron and Tibshirani(1993)}]{Efron:Tibshirani:1993}
Efron B, Tibshirani RJ (1993) An Introduction to the Bootstrap. Chapman \&
  Hall/CRC, New York

\bibitem[{Everitt and Hothorn(2009)}]{everitt09}
Everitt BS, Hothorn T (2009) A Handbook of Statistical Analyses Using {R},
  Second Edition. Chapman \& Hall, London

\bibitem[{Ford and Silvey(1980)}]{ford1980sequentially}
Ford I, Silvey S (1980) A sequentially constructed design for estimating a
  nonlinear parametric function. Biometrika 67(2):381--388

\bibitem[{Fraley and Raftery(1998)}]{fraley98b}
Fraley C, Raftery AE (1998) How many clusters? {W}hich clustering method?
  {A}nswers via model-based cluster analysis. Computer Journal 41:578--588

\bibitem[{Fraley and Raftery(2002)}]{fraley02}
Fraley C, Raftery AE (2002) Model-based clustering, discriminant analysis, and
  density estimation. Journal of the American Statistical Association
  97(458):611--612

\bibitem[{Fraley et~al(2012)Fraley, Raftery, Murphy, and Scrucca}]{fraley12}
Fraley C, Raftery AE, Murphy TB, Scrucca L (2012) mclust {V}ersion 4 for {R}:
  {N}ormal {M}ixture {M}odeling for {M}odel-{B}ased {C}lustering,
  {C}lassification, and {D}ensity {E}stimation. Tech. Rep. No. 597, Department
  of Statistics, University of Washington, USA.

\bibitem[{Gr{\"u}n and Leisch(2007)}]{grun2007fitting}
Gr{\"u}n B, Leisch F (2007) Fitting finite mixtures of generalized linear
  regressions in {R}. Computational Statistics \& Data Analysis
  51(11):5247--5252

\bibitem[{Hong et~al(2015)Hong, Mahajan, and Nekipelov}]{hong2015extremum}
Hong H, Mahajan A, Nekipelov D (2015) Extremum estimation and numerical
  derivatives. Journal of Econometrics 188(1):250--263

\bibitem[{Lee and McLachlan(2013{\natexlab{a}})}]{lee2013emmix}
Lee SX, McLachlan GJ (2013{\natexlab{a}}) {EMMIX}-uskew: an {R} package for
  fitting mixtures of multivariate skew t-distributions via the {EM} algorithm.
  Journal of Statistical Software 55(12):1--22

\bibitem[{Lee and McLachlan(2013{\natexlab{b}})}]{lee2013model}
Lee SX, McLachlan GJ (2013{\natexlab{b}}) Model-based clustering and
  classification with non-normal mixture distributions. Statistical Methods \&
  Applications 22(4):427--454

\bibitem[{Leeb and P{\"o}tscher(2005)}]{Leeb2005}
Leeb H, P{\"o}tscher BM (2005) {Model selection and inference: Facts and
  fiction}. Econometric Theory 21(1):21--59

\bibitem[{McLachlan(1987)}]{mclachlan87}
McLachlan G (1987) On bootstrapping the likelihood ratio test statistic for the
  number of components in a normal mixture. Journal of the Royal Statistical
  Society Series C (Applied) 36:318--324

\bibitem[{McLachlan et~al(1999)McLachlan, Peel, Basford, and
  Adams}]{mclachlan99}
McLachlan G, Peel D, Basford K, Adams P (1999) Fitting mixtures of normal and
  $t$-components. Journal of Statistical Software 4(2)

\bibitem[{McLachlan and Krishnan(1997)}]{mclachlan97}
McLachlan GJ, Krishnan T (1997) The {EM} algorithm and extensions. John Wiley
  \& Sons Inc., New York

\bibitem[{McLachlan and Peel(2000)}]{mclachlan00}
McLachlan GJ, Peel D (2000) Finite mixture models. John Wiley \& Sons Inc., New
  York

\bibitem[{Meilijson(1989)}]{meilijson89}
Meilijson I (1989) A fast improvement to the {EM} algorithm on its own terms.
  Journal of the Royal Statistical Society Series B (Methodological)
  51(1):127--138

\bibitem[{Meng and Rubin(1991)}]{meng91}
Meng X, Rubin D (1991) Using {EM} to obtain asymptotic variance-covariance
  matrices: the {SEM} algorithm. Journal of the American Statistical
  Association 86(416):899--909

\bibitem[{Meng and Rubin(1989)}]{meng89}
Meng XL, Rubin D (1989) Obtaining asymptotic variance-covariance matrices for
  missing-data problems using {EM}. In: Proceedings of the American Statistical
  Association (Statistical Computing Section), American Statistical
  Association, Alexandria, Virginia, pp 140--144

\bibitem[{Mita et~al(2012)Mita, Jiao, Kani, Tabuchi, and Hara}]{mita2012}
Mita N, Jiao J, Kani K, Tabuchi A, Hara H (2012) The parametric and
  non-parametric bootstrap resamplings for the visual acuity measurement.
  Kawasaki Journal of Medical Welfare 18:19--28

\bibitem[{Moulton and Zeger(1991)}]{moulton1991bootstrapping}
Moulton LH, Zeger SL (1991) Bootstrapping generalized linear models.
  Computational Statistics \& Data Analysis 11(1):53--63

\bibitem[{Newton and Raftery(1994)}]{newton94}
Newton MA, Raftery AE (1994) Approximate {B}ayesian inference with the weighted
  likelihood bootstrap. Journal of the Royal Statistical Society Series B
  (Methodological) 56(1):3--26

\bibitem[{Nyamundanda et~al(2010)Nyamundanda, Brennan, and
  Gormley}]{nyamundanda2010}
Nyamundanda G, Brennan L, Gormley I (2010) Probabilistic principal component
  analysis for metabolomic data. BMC Bioinformatics 11(1):571

\bibitem[{Pawitan(2000)}]{Pawitan2000}
Pawitan Y (2000) Computing empirical likelihood from the bootstrap. Statistics
  \& Probability Letters 47(4):337--345

\bibitem[{Peel(1998)}]{peel98}
Peel D (1998) Mixture model clustering and related topics. PhD thesis,
  University of Queensland, Brisbane

\bibitem[{Quenouille(1956)}]{quenouille56}
Quenouille M (1956) Notes on bias in estimation. Biometrika 43(2):343--348

\bibitem[{{R Core Team}(2017)}]{R2017}
{R Core Team} (2017) R: A Language and Environment for Statistical Computing. R
  Foundation for Statistical Computing, Vienna, Austria,
  \urlprefix\url{https://www.R-project.org/}

\bibitem[{Schwarz(1978)}]{schwarz1978}
Schwarz G (1978) Estimating the dimension of a model. Annals of Statistics
  6(2):461--464

\bibitem[{Shi(1988)}]{shi1988note}
Shi X (1988) A note on the delete-d jackknife variance estimators. Statistics
  \& Probability Letters 6(5):341--347

\bibitem[{Stoica and S{\"o}derstr{\"o}m(1982)}]{stoica1982non}
Stoica P, S{\"o}derstr{\"o}m T (1982) On non-singular information matrices and
  local identifiability. International Journal of Control 36(2):323--329

\bibitem[{Tanner(2012)}]{tanner_2012}
Tanner MA (2012) Tools for statistical inference. Springer, New York

\bibitem[{Titterington(1984)}]{titterington1984recursive}
Titterington DM (1984) Recursive parameter estimation using incomplete data.
  Journal of the Royal Statistical Society Series B (Methodological)
  46(2):257--267

\bibitem[{Tukey(1958)}]{tukey58}
Tukey J (1958) Bias and confidence in not-quite large samples (abstract). The
  Annals of Mathematical Statistics 29(2):614

\bibitem[{Turner(2000)}]{turner2000estimating}
Turner TR (2000) Estimating the propagation rate of a viral infection of potato
  plants via mixtures of regressions. Journal of the Royal Statistical Society:
  Series {C} (Applied Statistics) 49(3):371--384

\bibitem[{Wu(1986)}]{wu1986jackknife}
Wu CFJ (1986) Jackknife, bootstrap and other resampling methods in regression
  analysis. the Annals of Statistics 14(4):1261--1295

\end{thebibliography}

\newpage
\appendix
\section{Pairs plots of a simulated data set from \emph{Simulation Setting Three}.}
\label{app:sim3plots}

\begin{figure}[!h]
\begin{center}
\includegraphics[height=0.6\textheight, width=\textwidth, angle=270]{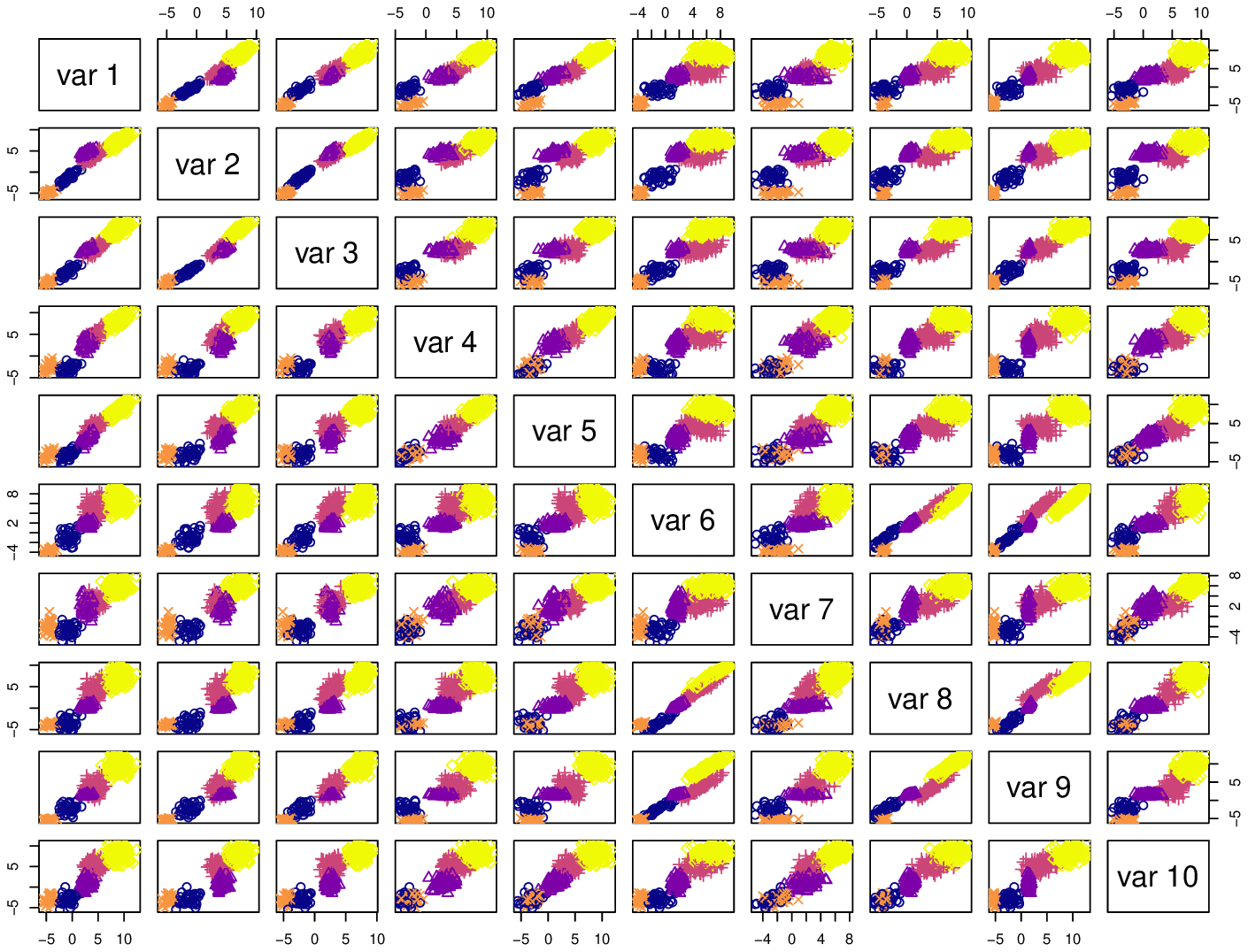}
\caption{Pairs plots of the first $10$ variables for a single simulated data set from \emph{Simulation Setting Three} ($n = 500, p = 25, G = 5$).}
\label{fig:M91}
\end{center}
\end{figure}

\begin{figure}[!h]
\begin{center}
\includegraphics[height=0.6\textheight, width=\textwidth, angle=270]{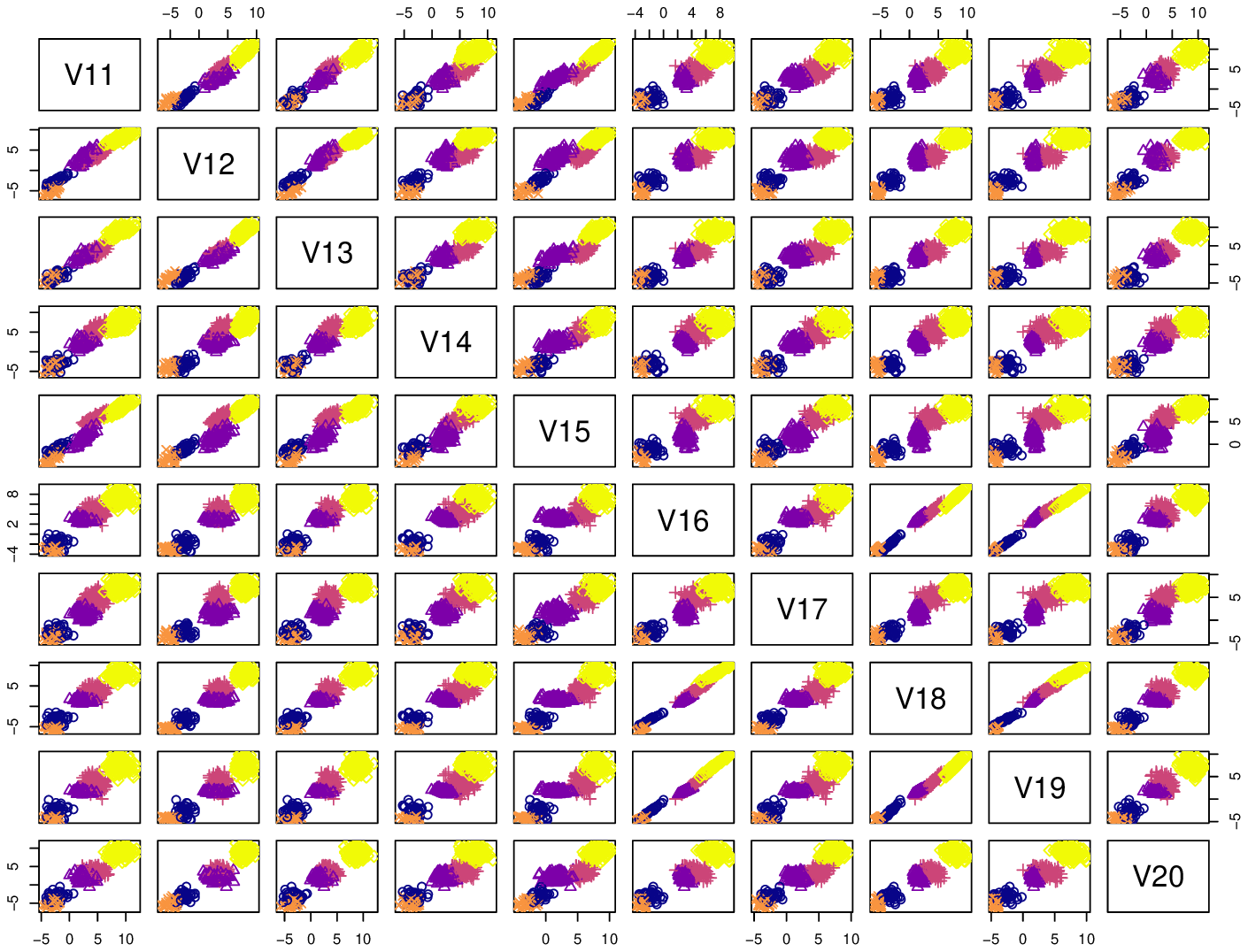}
\caption{Pairs plots of the second $10$ variables for a single simulated data set from \emph{Simulation Setting Three} ($n = 500, p = 25, G = 5$).}
\label{fig:M92}
\end{center}
\end{figure}

\begin{figure}[!h]
\begin{center}
\includegraphics[height=0.6\textheight, width=0.8\textwidth, angle=270]{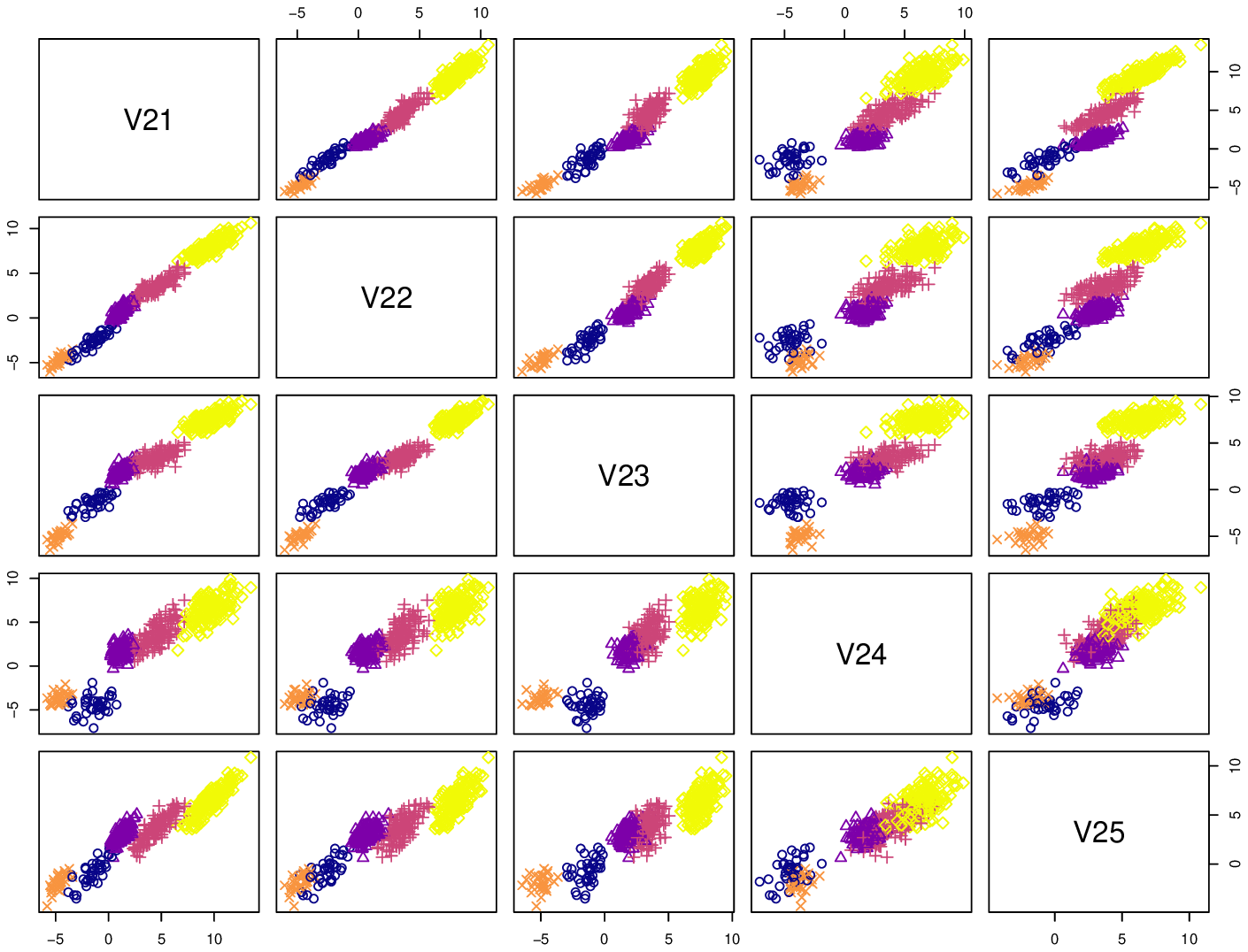}
\caption{Pairs plots of the final $5$ variables for a single simulated data set from \emph{Simulation Setting Three} ($n = 500, p = 25, G = 5$).}
\label{fig:M93}
\end{center}
\end{figure}

\emph{Simulation Setting Three} explores the performance and computational features of the JK, BS, PB and WLBS approaches to parameter variance estimation in a higher dimensional setting featuring overlapping and small clusters. Figures $\ref{fig:M91}$, $\ref{fig:M92}$ and $\ref{fig:M93}$ provide pairs plots from a single simulated data set under this setting for which $n = 500$, $p = 25$ and $G = 5$. Each of the different colours/symbols in the plots denotes one of the $5$ distinct clusters of observations simulated.

\newpage

\newpage

\newpage
\section{Covariance parameter estimates and standard errors for the \emph{Thyroid} data}
\label{app:thyroid}

\noindent
Cluster covariance estimated values are presented below using jackknife (JK), bootstrap (BS), parametric bootstrap (PB) and weighted likelihood bootstrap (WLBS) methods (with associated standard errors) for the optimal mixture of Gaussians model for the \emph{Thyroid} data, group $1$, where $G = 3$ and $p = 5$ and the optimal model has unequal diagonal covariance structure across clusters.

\begin{center}
\begin{tabular}{ll}
$\Sigma_{MCLUST, \,\,Group \,1}$ &
$= \left(\begin{array}{ccccc}
66.39 &       0    &       0    &       0    &       0  \\
0  &       4.82    &       0    &       0    &       0   \\
0  &       0    &       0.23  &       0  &       0 \\
0  &       0   &       0   &       0.22  &       0  \\
0  &       0    &       0    &       0    &       3.19   \\
\end{array}\right)$ \\
& \\
$\Sigma_{JK, \,\,Group \,1}$ &
$= \left(\begin{array}{ccccc}
67.50 \:(7.82) &       0   \:(0) &       0   \:(0) &       0   \:(0) &       0   \:(0)\\
0 \:(0) &       4.80   \:(0.63) &       0   \:(0) &       0   \:(0) &       0   \:(0)\\
0 \:(0) &       0   \:(0) &       0.24   \:(0.03) &       0   \:(0) &       0   \:(0)\\
0 \:(0) &       0   \:(0) &       0   \:(0) &       0.33   \:(0.04) &       0   \:(0)\\
0 \:(0) &       0   \:(0) &       0   \:(0) &       0   \:(0) &       3.25   \:(0.36)\\
\end{array}\right)$ \\
& \\
$\Sigma_{BS, \,\,Group \,1}$ &
$= \left(\begin{array}{ccccc}
66.00 \:(8.25) &       0   \:(0) &       0   \:(0) &       0   \:(0) &       0   \:(0)\\
0 \:(0) &       4.80   \:(0.64) &       0   \:(0) &       0   \:(0) &       0   \:(0)\\
0 \:(0) &       0   \:(0) &       0.23   \:(0.03) &       0   \:(0) &       0   \:(0)\\
0 \:(0) &       0   \:(0) &       0   \:(0) &       0.22   \:(0.05) &       0   \:(0)\\
0 \:(0) &       0   \:(0) &       0   \:(0) &       0   \:(0) &       3.16   \:(0.34)\\
\end{array}\right)$ \\
& \\
$\Sigma_{PB, \,\,Group \,1}$ &
$= \left(\begin{array}{ccccc}
65.85 \:(7.80) &       0   \:(0) &       0   \:(0) &       0   \:(0) &       0   \:(0)\\
0 \:(0) &       4.80   \:(0.54) &       0   \:(0) &       0   \:(0) &       0   \:(0)\\
0 \:(0) &       0   \:(0) &       0.23   \:(0.03) &       0   \:(0) &       0   \:(0)\\
0 \:(0) &       0   \:(0) &       0   \:(0) &       0.22   \:(0.03) &       0   \:(0)\\
0 \:(0) &       0   \:(0) &       0   \:(0) &       0   \:(0) &       3.17   \:(0.37)\\
\end{array}\right)$ \\
& \\
$\Sigma_{WLBS, \,\,Group \,1}$ &
$= \left(\begin{array}{ccccc}
65.85 \:(7.99) &       0   \:(0) &       0   \:(0) &       0   \:(0) &       0   \:(0)\\
0 \:(0) &       4.78   \:(0.62) &       0   \:(0) &       0   \:(0) &       0   \:(0)\\
0 \:(0) &       0   \:(0) &       0.23   \:(0.03) &       0   \:(0) &       0   \:(0)\\
0 \:(0) &       0   \:(0) &       0   \:(0) &       0.22   \:(0.05) &       0   \:(0)\\
0 \:(0) &       0   \:(0) &       0   \:(0) &       0   \:(0) &       3.17   \:(0.42)\\
\end{array}\right)$ \\
& \\
\end{tabular}
\end{center}

\vspace{5mm}
\noindent
Cluster covariance estimated values are presented below using jackknife (JK), bootstrap (BS), parametric bootstrap (PB) and weighted likelihood bootstrap (WLBS) methods (with associated standard errors) for the optimal mixture of Gaussians model for the \emph{Thyroid} data, group $2$, where $G = 3$ and $p = 5$ and the optimal model has unequal diagonal covariance structure across clusters.

\begin{center}
\begin{tabular}{ll}
$\Sigma_{MCLUST, \,\,Group \,2}$ &
$= \left(\begin{array}{ccccc}
344.46 &       0    &       0    &       0    &       0  \\
0  &       17.44    &       0    &       0    &       0   \\
0  &       0    &       4.92  &       0  &       0 \\
0  &       0   &       0   &       0.15  &       0  \\
0  &       0    &       0    &       0    &       0.07   \\
\end{array}\right)$ \\
& \\
$\Sigma_{JK, \,\,Group \,2}$ &
$= \left(\begin{array}{ccccc}
384.31 \:(101.72) &       0   \:(0) &       0   \:(0) &       0   \:(0) &       0   \:(0)\\
0 \:(0) &       14.84  \:(3.00) &       0   \:(0) &       0   \:(0) &       0   \:(0)\\
0 \:(0) &       0   \:(0) &       5.19 \:(1.37) &       0   \:(0) &       0   \:(0)\\
0 \:(0) &       0   \:(0) &       0   \:(0) &       0.15   \:(0.03) &       0   \:(0)\\
0 \:(0) &       0   \:(0) &       0   \:(0) &       0   \:(0) &       0.08   \:(0.02)\\
\end{array}\right)$ \\
& \\
$\Sigma_{BS, \,\,Group \,2}$ &
$= \left(\begin{array}{ccccc}
336.73 \:(98.03) &       0   \:(0) &       0   \:(0) &       0   \:(0) &       0   \:(0)\\
0 \:(0) &       16.85   \:(2.88) &       0   \:(0) &       0   \:(0) &       0   \:(0)\\
0 \:(0) &       0   \:(0) &       4.77   \:(1.31) &       0   \:(0) &       0   \:(0)\\
0 \:(0) &       0   \:(0) &       0   \:(0) &       0.15   \:(0.03) &       0   \:(0)\\
0 \:(0) &       0   \:(0) &       0   \:(0) &       0   \:(0) &       0.07   \:(0.02)\\
\end{array}\right)$ \\
& \\
$\Sigma_{PB, \,\,Group \,2}$ &
$= \left(\begin{array}{ccccc}
334.34 \:(83.28) &       0   \:(0) &       0   \:(0) &       0   \:(0) &       0   \:(0)\\
0 \:(0) &       17.04   \:(4.45) &       0   \:(0) &       0   \:(0) &       0   \:(0)\\
0 \:(0) &       0   \:(0) &       4.74   \:(1.15) &       0   \:(0) &       0   \:(0)\\
0 \:(0) &       0   \:(0) &       0   \:(0) &       0.15   \:(0.04) &       0   \:(0)\\
0 \:(0) &       0   \:(0) &       0   \:(0) &       0   \:(0) &       0.07   \:(0.02)\\
\end{array}\right)$ \\
& \\
$\Sigma_{WLBS, \,\,Group \,2}$ &
$= \left(\begin{array}{ccccc}
332.50 \:(92.04) &       0   \:(0) &       0   \:(0) &       0   \:(0) &       0   \:(0)\\
0 \:(0) &       16.71  \:(2.71) &       0   \:(0) &       0   \:(0) &       0   \:(0)\\
0 \:(0) &       0   \:(0) &       4.81   \:(1.28) &       0   \:(0) &       0   \:(0)\\
0 \:(0) &       0   \:(0) &       0   \:(0) &       0.15   \:(0.03) &       0   \:(0)\\
0 \:(0) &       0   \:(0) &       0   \:(0) &       0   \:(0) &       0.07   \:(0.02)\\
\end{array}\right)$ \\
& \\
\end{tabular}
\end{center}

\vspace{5mm}
\noindent
Cluster covariance estimated values are presented below using jackknife (JK), bootstrap (BS), parametric bootstrap (PB) and weighted likelihood bootstrap (WLBS) methods (with associated standard errors) for the optimal mixture of Gaussians model for the \emph{Thyroid} data, group $3$, where $G = 3$ and $p = 5$ and the optimal model has unequal diagonal covariance structure across clusters.

\begin{center}
\begin{tabular}{ll}
$\Sigma_{MCLUST, \,\,Group \,3}$ &
$= \left(\begin{array}{ccccc}
95.23 &       0    &       0    &       0    &       0  \\
0  &       4.26    &       0    &       0    &       0   \\
0  &       0    &       0.28  &       0  &       0 \\
0  &       0   &       0   &       147.06  &       0  \\
0  &       0    &       0    &       0    &       231.22   \\
\end{array}\right)$ \\
& \\
$\Sigma_{JK, \,\,Group \,3}$ &
$= \left(\begin{array}{ccccc}
95.47 \:(29.87) &       0   \:(0) &       0   \:(0) &       0   \:(0) &       0   \:(0)\\
0 \:(0) &       2.91  \:(1.10) &       0   \:(0) &       0   \:(0) &       0   \:(0)\\
0 \:(0) &       0   \:(0) &       0.24   \:(0.06) &       0   \:(0) &       0   \:(0)\\
0 \:(0) &       0   \:(0) &       0   \:(0) &       157.52   \:(71.60) &       0   \:(0)\\
0 \:(0) &       0   \:(0) &       0   \:(0) &       0   \:(0) &       234.45   \:(71.18)\\
\end{array}\right)$ \\
& \\
$\Sigma_{BS, \,\,Group \,3}$ &
$= \left(\begin{array}{ccccc}
90.83 \:(27.53) &       0   \:(0) &       0   \:(0) &       0   \:(0) &       0   \:(0)\\
0 \:(0) &       3.93   \:(0.94) &       0   \:(0) &       0   \:(0) &       0   \:(0)\\
0 \:(0) &       0   \:(0) &       0.26   \:(0.06) &       0   \:(0) &       0   \:(0)\\
0 \:(0) &       0   \:(0) &       0   \:(0) &       143.33   \:(65.03) &       0   \:(0)\\
0 \:(0) &       0   \:(0) &       0   \:(0) &       0   \:(0) &       222.37   \:(65.83)\\
\end{array}\right)$ \\
& \\
$\Sigma_{PB, \,\,Group \,3}$ &
$= \left(\begin{array}{ccccc}
91.11 \:(25.03) &       0   \:(0) &       0   \:(0) &       0   \:(0) &       0   \:(0)\\
0 \:(0) &       4.17   \:(1.16) &       0   \:(0) &       0   \:(0) &       0   \:(0)\\
0 \:(0) &       0   \:(0) &       0.27   \:(0.08) &       0   \:(0) &       0   \:(0)\\
0 \:(0) &       0   \:(0) &       0   \:(0) &       141.92   \:(38.84) &       0   \:(0)\\
0 \:(0) &       0   \:(0) &       0   \:(0) &       0   \:(0) &       222.99   \:(61.33)\\
\end{array}\right)$ \\
& \\
$\Sigma_{WLBS, \,\,Group \,3}$ &
$= \left(\begin{array}{ccccc}
92.72 \:(25.66) &       0   \:(0) &       0   \:(0) &       0   \:(0) &       0   \:(0)\\
0 \:(0) &       3.91   \:(0.85) &       0   \:(0) &       0   \:(0) &       0   \:(0)\\
0 \:(0) &       0   \:(0) &       0.26   \:(0.05) &       0   \:(0) &       0   \:(0)\\
0 \:(0) &       0   \:(0) &       0   \:(0) &       139.92   \:(61.20) &       0   \:(0)\\
0 \:(0) &       0   \:(0) &       0   \:(0) &       0   \:(0) &       219.38   \:(62.58)\\
\end{array}\right)$ \\
& \\
\end{tabular}
\end{center}

\vspace{5mm}
\noindent
The following code produces all variance estimation results for the \emph{Thyroid} data set, using the \texttt{MclustBootstrap} function in \texttt{mclust}.

\begin{center}
\begin{verbatim}
library(mclust)
data(thyroid)
object = Mclust(thyroid[,2:6], G = 3)
jack = MclustBootstrap(object, type = "jk")
boot = MclustBootstrap(object, type = "bs")
pb = MclustBootstrap(object, type = "pb")
wlbs = MclustBootstrap(object, type = "wlbs")
summary(jack, what = "se")
summary(boot, what = "se")
summary(pb, what="se")
summary(wlbs, what = "se")
\end{verbatim}
\end{center}

\end{document}